\documentclass[12pt,draftclsnofoot,onecolumn]{IEEEtran}
\ifCLASSINFOpdf
   \usepackage[pdftex]{graphicx}
   \graphicspath{{../pdf/}{../jpeg/}}
   \DeclareGraphicsExtensions{.pdf,.jpeg,.png}
\else
   \usepackage[dvips]{graphicx}
    \usepackage{subfigure}
   \graphicspath{{../eps/}}
   \DeclareGraphicsExtensions{.eps}
\fi
\usepackage{cite}

\usepackage{makecell}

\usepackage{subfigure}
\usepackage{latexsym,bm}
\usepackage{amssymb,amsfonts,amsbsy,amscd,amsgen,amsopn,amsxtra}
\usepackage{multirow}

\usepackage{algorithmic} 
\usepackage[ruled,vlined]{algorithm2e}
\usepackage{array}
\makeatletter
\newcommand{\thickhline}{%
    \noalign {\ifnum 0=`}\fi \hrule height 0.75pt
    \futurelet \reserved@a \@xhline
}
\newcolumntype{"}{@{\hskip\tabcolsep\vrule width 1pt\hskip\tabcolsep}}
\makeatother
\usepackage{amsmath}
\usepackage{xcolor}
\usepackage{arydshln}
\usepackage{slashbox}
\usepackage{calc}
\usepackage{tikz}

\usepackage{amsthm}
\theoremstyle{definition}

\theoremstyle{definition}

\ifCLASSINFOpdf

\else

\fi

\begin{document}

\title{Enumeration of Minimum Hamming Weight Polar Codewords with Sublinear Complexity}

\author{Fengyi Cheng,
        Aijun Liu,
        Jincheng Dai, Kai Niu, and Xiaohu Liang
        \thanks{This work is supported by the National Natural Science Foundation of
China (No.61501508, No.61671476 and No.61901516), in part by the Natural
Science Foundation of Jiangsu Province of China (No.BK20180578) and China Postdoctoral Science Foundation (No.2019M651648). Fengyi Cheng, Aijun Liu and Xiaohu Liang are with the department of Communication Engineering,  Army Engineering University of PLA, Nanjing 210007, China (e-mail: tech$\_$cfy@126.com, liuaj.cn@163.com, liangxiaohu688@163.com). Jincheng Dai and Kai Niu are with the Key Laboratory of Universal Wireless Communications, Ministry of Education, Beijing University of Posts and Telecommunications (BUPT), Beijing 100876, China (email: daijincheng@bupt.edu.cn, niukai@bupt.edu.cn)}
        \thanks{    }
\vspace{-1em}

}

\markboth{}%
{Cheng \MakeLowercase{\textit{et al.}}: Enumeration of Minimum Hamming Weight Polar Codewords with Sublinear Complexity}
\maketitle

\begin{abstract}
Polar code, with explicit construction and recursive structure, is the latest breakthrough in channel coding field for its low-complexity and theoretically capacity-achieving property. Since polar codes can approach the maximum likelihood performance under successive cancellation list decoding (SCLD), its decoding performance can be evaluated by Bonferroni-type bounds (e.g., union bound) in which the Hamming weight spectrum will be used. Especially, the polar codewords with minimum Hamming weight (PC-MHW) are the most important item in that bound because they  make major contributions to the decoding error pattern particularly at high signal-to-noise-ratio. In this work, we propose an efficient strategy for enumerating the PC-MHW and its number. By reviewing the inherent reason that PC-MHW can be generated by SCLD, we obtain some common features of PC-MHW captured by SCLD. Using these features, we introduce a concept of \emph{zero-capacity bit-channel} to obtain a tight upper bound for the number of PC-MHW, whose computing complexity is sublinear with code length. Furthermore, we prove that the proposed upper bound is really the exact number of PC-MHW in most cases. Guided by the bound and its theoretical analysis, we devise an efficient SCLD-based method to enumerate PC-MHW, which requires less than half of the list size compared with the existing methods.
\end{abstract}


%
\IEEEpeerreviewmaketitle

\section{Introduction}
%
%
%
%

\IEEEPARstart{P}OLAR codes can achieve Shannon capacity under successive cancellation decoding (SCD) as the code length goes to infinity \cite{original}. For the moderate code length, however, SC decoding (SCD) can not provide satisfactory performance. To overcome this shortcoming,  successive cancellation list decoding (SCLD) was proposed by \cite{original_SCL}. Unlike SCD, SCLD can reserve $L$ most reliable decoding paths, where the reliability is evaluated by a path metric (PM). After that, some algorithms are proposed to improve the decoding latency, memory space and power overhead of SCLD \cite{LLRSCL}-\cite{enhanceSCL2}. Especially, assisted by cyclic redundancy check (CRC) \cite{original_SCL} \cite{CRC}, the performance of SCLD can be further improved to make polar codes as the coding scheme for the control channel in the $5^{th}$ generation wireless communication standards \cite{3G}.

Considering SCLD can approach the ML performance even for practical list size $L$ (e.g., $L{\rm{=}}8$), its performance, especially at high signal to noise ratio (SNR), can be evaluated by the union bound where the number of polar codewords with minimum hamming weight (PC-MHW) is the most important item \cite{MLbound}. Moreover, \cite{Ada} further verifies that the inherent reason that
CRC can improve the performance of SCLD is preventing the erroneously decoded minimum
Hamming distance codewords from passing the check.  The first method to enumerate PC-MHW and its number is proposed in \cite{searchMHWD}, where the authors verify that if the all-zero codeword is BPSK modulated and transmitted by the noiseless AWGN channel, then the information vector whose corresponding codeword has MHW will survive in the $L$ remaining paths on the competition of SCLD. Based on the one to one relationship between the codeword with its information vector, the PC-MHW can also be seen as the output of SCLD. However, this method requires large computation complexity and memory space. To solve this problem, \cite{CRC_design} proposes a searching strategy which can divide the set of PC-MHW into several subsets for searching, so as to narrow down the required list size of SCLD. In practical, both the used list size in these two methods should be set larger than the actually required value to prevent omission. This extra list size, which can be regarded as unnecessary overhead, can be avoided if the number of PC-MHW can be approximately predicted before searching. In \cite{ProCom}, a probabilistic computation method is proposed to evaluate the Hamming weight spectrum of polar codes with complexity $O(N^5)$, where $N$ is the code length. Thereafter, the accuracy of this evaluation is enhanced by \cite{enhanceProCom} and the complexity can be reduced to $O(N^3)$. However, this method is tenable only at high code rate. Moreover, the complexity of these two methods are still too high to analyze the codes with moderate or long code length.

In this paper, we propose an efficient strategy to enumerate the PC-MHW and its number. Based on the fact that the PC-MHW can be searched by SCLD, we analyze the characters of the PC-MHW searched by the SCLD. Then, guided by these characters, we propose a tight upper bound for evaluating the number of PC-MHW with sublinear complexity $O(\log_2N)$.  Based on such bound, we further propose an efficient strategy to search the PC-MHW by less than half of the list size required in the existing methods.

The highlights of our contributions are summarized as follows:
\begin{enumerate}
  \item The essential reason that the PC-MHW can be generated by SCLD is reviewed. For each path of SCLD, there may exist some bits that does not be hard decided according to the corresponding decoding log-likelihood ratio (LLR). The set of the locations of such bits is referred to as reverse decision set (RDS). We prove that the set of PC-MHW can be divided into several subsets to ensure  the information vectors of the PC-MHW in a same subset share the same RDS when they are taken as the output of SCLD-based searching.
  \item We give a tight upper bound of the number of PC-MHW. For each subset of PC-MHW, divided according to their RDS obtained in  SCLD-based searching, we can give an upper bound for its cardinality. By adding all such upper bounds, the bound for the total number of the PC-MHW can be derived. We further demonstrate that this upper bound, obtained with sublinear complexity $O(\log_2N)$, is really the exact number of PC-MHW at most code rate or code length.
  \item We propose an efficient strategy for enumerating PC-MHW. Guided by the proposed upper bound and the theoretical analysis of it, we can further divide the subset of PC-MHW into several smaller subsets. In once searching, we only need to search one smaller subset but instead of the whole PC-MHW. Thus, the required list size of SCLD used for searching can be further narrowed down to less than half of that used in the existing methods, so as to reduce the complexity and space memory for searching.
\end{enumerate}

The reminder of this paper is organized as follows. Section II describes some basic conception associated with polar codes, which will be incurred in the following paper. In Section III, we will reveal how the information vectors of PC-MHW can be enumerated by SCLD so as to obtain their common features captured by SCLD. An upper bound of the number of PC-MHW is described in Section IV. In Section V, an efficient strategy for enumerating PC-MHW is proposed. Simulation results are given in Section VI, and conclusions are drawn in Section VII.

By necessity, this paper contains a fair amount of theoretical proof. Thus, on a first reading, the reader is advised to preview the Section III.A which will provide a high-level description of the proposed enumeration for the number of PC-MHW.
\section{PRELIMINARIES}
\subsection{Notation Conventions}
In this paper, we use lowercase letters, such as $x$, to denote scalars. $\left\lceil x \right\rceil $ is a ceiling function of a float value $x$. We write calligraphic characters (e.g., $\mathcal{X}$) to denote sets. $|\mathcal{X}|$ is cardinality of $\mathcal{X}$. $\mathcal{X}{\rm{-}}\mathcal{Y}$ means the difference set between $\mathcal{X}$ and $\mathcal{Y}$. $\phi$ stands for null vector or null set. The notation $x_i^j$, with $i{\rm{<}}j$, is used to denote a vector [$x_i$,$x_{i+1}$,...,$x_{j}$]. If $i{\rm{>}}j$, $x_i^j{\rm{=}}\phi$. If $i{\rm{=}}j$, $x_i^j{\rm{=}}x_i$. When the dimension does not need to be emphasized, we also use bold lowercase letters or Greek letters, such as $\mathbf{x}$ or $\alpha$, to denote vectors. $\mathbf{0}_i$ and $\mathbf{1}_i$ stand for $i$-length all-zero and all-one vector, respectively. Note that when $i{\rm{\leq}}0$, $\mathbf{0}_i$ and $\mathbf{1}_i$ are both null vector. Let ${\left[\kern-0.15em\left[ {i,j}
 \right]\kern-0.15em\right]}$ be the set of consecutive integer $\{i,i{\rm{+}}1,...,j\}$. Given an index set $\mathcal{X}{\rm{\subseteq}}{\left[\kern-0.15em\left[ {i,j}
 \right]\kern-0.15em\right]}$, let $\mathbf{u}_\mathcal{X}$ denote the subvector of $u_i^j$, which consists of $u_k$s with $k{\rm{\in}}\mathcal{X}$. $w(\mathbf{u})$ denotes the Hamming weight of $\mathbf{u}$. Bold letters, such as $\mathbf{X}$, denote matrices. $\mathcal{B}$ and $\mathcal{Z}^*$ denote the set of binary $\{0,1\}$ and positive integer, respectively. $\otimes$ is the Kronecker product. $\textit{N}(a,b)$ denotes Gaussian distribution with mean $a$ and variance $b$. $\rm{sign}(\cdot)$ is the sign function. $\min(\mathcal{X})$ is the minimum value in set $\mathcal{X}$.

Let $\mathbf{b}_n(i){\rm{=}}[b_{i,1},b_{i,2},...,b_{i,n}]$ be the vector of binary expansion of integer $i$,  where the $b_{i,1}$ is the least significant bit. $\mathbf{b}_n(i)_j^k$ stands for vector $[b_{i,j},b_{i,j+1},...,b_{i,k}]$, with $1\leq j\leq k\leq n$.

For vector $\mathbf{x}$ that contains element $\alpha$, $\mathcal{P}_\alpha[\mathbf{x}]$ denotes the set of positions of $\alpha$ in $\mathbf{x}$. $\mathcal{P}_\alpha[\mathbf{x}]_j$ represents an element in $\mathcal{P}_\alpha[\mathbf{x}]$,  $j{\rm{=}}1,2,...,|\mathcal{P}_\alpha[\mathbf{x}]|$, and we assume  $\mathcal{P}_\alpha[\mathbf{x}]_1{\rm{<}}\mathcal{P}_\alpha[\mathbf{x}]_2{\rm{<}}...{\rm{<}}\mathcal{P}_\alpha[\mathbf{x}]_{|\mathcal{P}_\alpha[\mathbf{x}]|}$.

\emph{Example 1:} For $\mathbf{x}=[1,1,0,1,0]$, $\mathcal{P}_0[\mathbf{x}]{\rm{=}}\{3,5\}$,  $\mathcal{P}_0[\mathbf{x}]_1{\rm{=}}3$, $\mathcal{P}_0[\mathbf{x}]_2{\rm{=}}5$; $\mathcal{P}_1[\mathbf{x}]{\rm{=}}\{1,2,4\}$, $\mathcal{P}_1[\mathbf{x}]_1{\rm{=}}1$, $\mathcal{P}_1[\mathbf{x}]_2{\rm{=}}2$, $\mathcal{P}_1[\mathbf{x}]_3{\rm{=}}4$.
\subsection{Polar Codes}
A polar code with message length $K$ and code length $N{\rm{=}}2^n$, $n{\rm{=}}1$,$2$,$...$, is determined by  matrix $\mathbf{G}_N{\rm{=}}\tiny{{\left[\begin{array}{l}
1\;0\\
1\;1
\end{array} \right]}^{ \otimes n}}$ and the information set $\mathcal{A}{\rm{\subseteq}}\left[\kern-0.15em\left[ {1,N}
 \right]\kern-0.15em\right]$. Note that $\left| {{\mathcal{A}}} \right|{\rm{=}}K$ and the code rate $R$ is equal to $\frac{K}{N}$. Let $u_1^N$ be the information vector of polar code and $\mathbf{u}_\mathcal{A}$ be the vector of source message bits which are sent through the bit-channels with indices in $\mathcal{A}$. Let ${\mathcal{A}^c}{\rm{=}}\left[\kern-0.15em\left[ {1,N}
 \right]\kern-0.15em\right]{\rm{-}}\mathcal{A}$. The bits sent by the bit-channels with indices in $\mathcal{A}^c$ are fixed to 0. Hence, the encoding process of polar code can be expressed as $c_1^N=u_1^N\mathbf{G}_N=\mathbf{u}_\mathcal{A}\mathbf{G}_N^\mathcal{A}$,
where $c_1^N$ is the codeword and $\mathbf{G}_N^\mathcal{A}$ is the generator matrix which is composed of rows in $\mathbf{G}_N$ with indices in $\mathcal{A}$. The MHW of polar codes with $\mathbf{G}_N^\mathcal{A}$ is denoted by $d_m$. Throughout this paper, we assume that the codeword is BPSK modulated by  $\{0,1\}\rightarrow\{1,-1\}$. Thus, the received sequence $y_1^N$ satisfies $y_i=(1-2c_i)+n_i$, where $n_i\thicksim\textit{{N}}(0,\sigma^2)$ is an additive white Gaussian noise (AWGN).
\subsection{Successive-Cancellation Decoding}
The process of SC decoding can be depicted on a code tree as shown in Fig.1(a) \cite{SCcodetree}. The stage of the root node is the \emph{depth} of the tree. For $N$-length polar code, the code tree is composed of $2N{\rm{-}}1$ nodes and $n{\rm{+}}1$ stages.  The root node and the leaf nodes are at the stage $n$ and $0$, respectively. Let $V_\lambda^{(j)}$ denote the $j$-th node (counting from the left) at stage $\lambda$, with $j{\rm{\in}}\left[\kern-0.25em\left[ {1,2^{N-\lambda}}
 \right]\kern-0.25em\right]$. Each node has a LLR vector $\alpha[V_\lambda^{(j)}]$ and a codeword vector $\beta[V_\lambda^{(j)}]$. Both of these two message vectors are $2^\lambda$-length and they can be written as
\begin{equation}
\alpha[V_\lambda^{(j)}]{\rm{=}}\alpha[V_\lambda^{(j)}]_1^{2^\lambda}{\rm{=}}\left[{\alpha[V_\lambda^{(j)}]_1,\alpha[V_\lambda^{(j)}]_2,...,\alpha[V_\lambda^{(j)}]_{2^\lambda}} \right]
\end{equation}
\begin{equation}
\beta[V_\lambda^{(j)}]{\rm{=}}\beta[V_\lambda^{(j)}]_1^{2^\lambda}{\rm{=}}\left[{\beta[V_\lambda^{(j)}]_1,\beta[V_\lambda^{(j)}]_2,...,\beta[V_\lambda^{(j)}]_{2^\lambda}} \right]
\end{equation}
Actually, the SC decoding process is the process of calculating such two types of message vectors for each node. The LLR vectors are calculated from stage $n$ to stage $0$ while the codeword vectors are updated from stage $0$ to stage $n$.

\emph{Definition 1 (decoding LLR, decoding bit and input vector): For any leaf node $V_0^{(j)}$, $j{\rm{\in}}\left[\kern-0.15em\left[ {1,N}
 \right]\kern-0.15em\right]$, $\alpha[V_0^{(j)}]$ and $\beta[V_0^{(j)}]$ have only one element. $\alpha[V_0^{(j)}]$ is referred to as decoding LLR. $\beta[V_0^{(j)}]$ is called the decoding bit. The LLR vector of the root node is called input vector of decoder.}

At stage $n$, the input vector of decoder is set by $\alpha[V_n^{(1)}]=y_1^N$.

For any $\lambda{\rm{\in}}\left[\kern-0.15em\left[ {1,n{\rm{-}}1}
 \right]\kern-0.15em\right]$ and $j{\rm{\in}}\left[\kern-0.25em\left[ {1,2^\lambda}
 \right]\kern-0.25em\right]$, node $V_\lambda^{(j)}$ has a parent node (${{V}}_{\lambda  + 1}^{\left( {\left\lceil {{j \mathord{\left/
 {\vphantom {j 2}} \right.
 \kern-\nulldelimiterspace} 2}} \right\rceil } \right)}$), a left child node (${{V}}_{\lambda  - 1}^{\left( {2j - 1} \right)}$) and a right child node (${{V}}_{\lambda  - 1}^{\left( {2j} \right)}$). As shown in Fig.1(b),  $V_\lambda^{(j)}$ will participate in calculating the LLR vectors of its two child nodes (i.e., $\alpha[{{V}}_{\lambda  - 1}^{\left( {2j} \right)}]$ and $\alpha[{{V}}_{\lambda  - 1}^{\left( {2j - 1} \right)}]$) and send itself codeword vector $\beta[V_\lambda^{(j)}]$ to its parent node. When $\alpha[{V_\lambda^{(j)}}]$ is determined,  $V_\lambda^{(j)}$ is immediately activated to calculate $\alpha[{{V}}_{\lambda  - 1}^{\left( {2j - 1} \right)}]$ by
\begin{equation}\label{LLRnobit}
\alpha{[ {{{V}}_{\lambda  - 1}^{\left( {2j - 1} \right)}} ]_i} {\rm{=}} {\alpha{{[ {{{V}}_\lambda ^{\left( j \right)}} ]}_i}}{\rm{\boxplus}}{\alpha{{[ {{{V}}_\lambda ^{\left( j \right)}} ]}_{i + {2^{\lambda  - 1}}}}},\;\;i {\rm{\in}}\left[\kern-0.25em\left[ {1,2^{\lambda-1}}
 \right]\kern-0.25em\right]
\end{equation}
with $x{\rm{\boxplus}}y{\rm{:=}}{\rm{sign}}\left( x \right){\rm{sign}}\left( y \right)\min (\left\{ {\left| x \right|,\left| y \right|} \right\})$. After receiving $\beta[ {{{V}}_{\lambda  - 1}^{\left( {2j - 1} \right)}} ]$ from its child node, $V_\lambda^{(j)}$ will update $\alpha[ {{{V}}_{\lambda  - 1}^{\left( {2j} \right)}} ]$ by
\begin{equation}\label{LLRbit}
\alpha{[ {{{V}}_{\lambda  - 1}^{\left( {2j} \right)}} ]_i} {\rm{=}} \left( {1 {\rm{-}} 2\beta{{[ {{{V}}_{\lambda  - 1}^{\left( {2j - 1} \right)}} ]}_i}} \right)\alpha{[ {{{V}}_\lambda ^{\left( j \right)}} ]_i} {\rm{+}} \alpha{[ {{{V}}_\lambda ^{\left( j \right)}} ]_{i + {2^{\lambda-1} }}}
\end{equation}
with $i {\rm{\in}}\left[\kern-0.20em\left[ {1,2^{\lambda-1}}
 \right]\kern-0.20em\right]$. Then, $V_\lambda^{(j)}$ waits until it receives $\beta[{{V}}_{\lambda  - 1}^{\left( {2j} \right)}]$ and  $\beta[{{V}}_{\lambda  - 1}^{\left( {2j-1} \right)}]$ to update $\beta[V_\lambda^{(j)}]$
 \begin{equation}\label{bit}
\beta{[ {{{V}}_\lambda ^{\left( j \right)}} ]_i} = \left\{ \begin{gathered}
  \beta{[ {{{V}}_{\lambda  - 1}^{\left( {2j} \right)}} ]_i} \oplus \beta{[ {{{V}}_{\lambda  - 1}^{\left( {2j - 1} \right)}} ]_i},\;i \leqslant {2^{\lambda  - 1}} \hfill \\
  \beta{[ {{{V}}_{\lambda  - 1}^{\left( {2j} \right)}} ]_i},\quad \quad \quad \quad \quad \,\,\;\;\;i > {2^{\lambda  - 1}} \hfill \\
\end{gathered}  \right.
\end{equation}
If $\lambda{\rm{=}}0$, $\beta[V_\lambda^{(j)}]$ is the decoding bit $\hat u_j$ and can be derived by making hard decision of the decoding LLR $\alpha[V_0^{(j)}]$, i.e.,
 \begin{equation}\label{HBD}
{\hat u_j} {\rm{=}}\beta[V_0^{(j)}]{\rm{=}}h(\alpha[V_0^{(j)}]){\rm{=}} \left\{ \begin{gathered}
  0, \; j \in {{\mathcal{A}^c}}\;\operatorname{or} \;\alpha[V_0^{(j)}] > 0 \hfill \\
  1, \;\operatorname{others}  \hfill \\
\end{gathered}  \right.
\end{equation}

$\beta{[ {{{V}}_\lambda ^{\left( j \right)}} ]}$ can also be directly calculated  by the decoding bits
\begin{equation}\label{debit}
\beta{[ {{{V}}_\lambda ^{\left( j \right)}} ]}=\hat u_{(j-1){2^\lambda}+1}^{j{2^\lambda}}\mathbf{G}_{2^\lambda}
\end{equation}
For any $i{\rm{\in}}\left[\kern-0.25em\left[ {{(j{\rm{-}}1){2^\lambda}{\rm{+}}1},{j{2^\lambda}}}
 \right]\kern-0.25em\right]$, we can say leaf node $ {{{V}}_0 ^{\left( i \right)}}$ or decoding bit $\hat u_i$ participates in the calculating of $\beta{[ {{{V}}_\lambda ^{\left( j \right)}} ]}$. On completion of updating $\beta[V_\lambda^{(j)}]$, the operations associated with $V_\lambda^{(j)}$ are terminated and $V_\lambda^{(j)}$ will never be activated.
  \begin{figure}
\centering{\includegraphics[height=3cm]{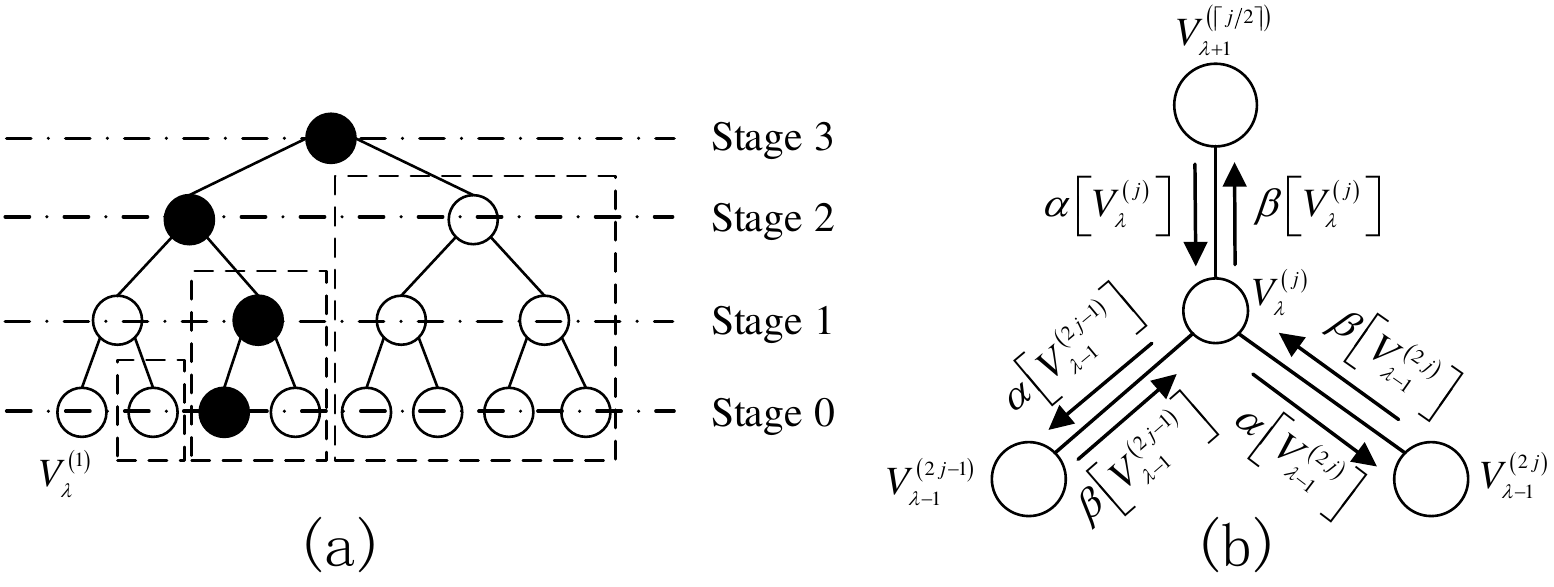}}
\caption{(a) A code tree with $N=8$. (b) Basic message processing element of code-tree with node $V_\lambda^{(j)}$.}
\end{figure}
\subsection{Successive-Cancellation List Decoding}
SCLD will reserve more than one decoding candidates (or paths). For any position $i{\rm{\in}}\mathcal{A}$, SCLD splits every decoding paths into 2 threads to consider both the probability of the current bits being 0 or 1. Thus, the decoding bits in one path may not be determined according to the suggestion of its LLR. To avoid an exponentially growing complexity, at most $L$ most reliable paths (or trajectories) could be reserved in the whole decoding process. Unlike SCD, when referring to a decoding trajectory of the SCLD, we should indicate the decoding step and the order of the path in the reserving list. The  $i$-th decoding step means that when $u_i$ is just decoded by SCLD. Let $u_i[l]_1^j{\rm{=}}[u_i[l]_1,u_i[l]_2,...,u_i[l]_j]$ denote the vector of the first $j$ bits of the $l$-th decoding path at the $i$-th decoding step, $i{\rm{>}}j$. Note that $u_i[l]_1^j$ is not necessarily equal to $u_{i+1}[l]_1^j$. Especially, if $i{\rm{<}}j$, $u_i[l]_1^j{\rm{=}}\phi$. When all the bits are decoded, the most reliable path will be output as the decoding result. The reliability is evaluated by a path metric (PM). In this paper, we adopt the LLR-based SCLD proposed in \cite{LLRSCL} whose PM is calculated in logarithmic domain. Concretely, the PM of decoding paths in this paper is defined as follows:

\emph{Definition 2 (PM and RDS): For any $i{\rm{\in}}\left[\kern-0.15em\left[ {1,N}
 \right]\kern-0.15em\right]$, the PM of the $l$-th decoded path $\hat{u}_i[l]_1^i$, $l{\rm{\in}}\left[\kern-0.15em\left[ {1,L}
 \right]\kern-0.15em\right]$, is defined by:}
\begin{equation}\label{PM}
\textsf{PM}[l]_i= {\tiny{\sum\nolimits_{j \in \overline{\mathcal{A}}_i[l]_1^i}}} {\left| L_i[l]_j\right|},
\end{equation}
\emph{where  ${L_i}{\left[ l \right]_j} {\rm{=}} \ln \frac{{\Pr \left( {{u_j} = 0\left| {y_1^N,{{\hat u}_i}\left[ l \right]_1^{j - 1}} \right.} \right)}}{{\Pr \left( {{u_j} = 1\left| {y_1^N,{{\hat u}_i}\left[ l \right]_1^{j - 1}} \right.} \right)}}$ is the decoding LLR for $\hat u_i[l]_j$, and $\overline{\mathcal{A}}_i[l]_1^i$, called reverse decision set (RDS) of path $\hat u_i[l]_1^i$, is the set of positions at which  $\hat u_i[l]_1^i$ does not make hard decision based on decoding LLR.}

Note that RDS can also include the elements in $\mathcal{A}^c$. In the following, we simply use $\overline{\mathcal{A}}_N[l]$ to denote the RDS of $\hat u_N[l]_1^N$.
\subsection{Searching for PC-MHW}
For polar codes with $\mathbf{G}_N^{\mathcal{A}}$, let $\mathcal{U}_{N,m}$ be the set of information vector whose corresponding codeword has MHW. $d_m$ is equal to the minimum row weight of $\mathbf{G}_N^{\mathcal{A}}$ \cite{MHD}. $\mathcal{U}_{N,m}$ can be searched by SCLD under the condition that all-0 polar codeword is BPSK modulated and sent by noiseless AWGN channel \cite{searchMHWD}. In the following paper, this condition will be equivalently defined as
\begin{equation}
y_1^N=\mathbf{1}_N.
\end{equation}
Under this condition, on completion of SCLD, if we first discard the all-0 path, then, from the remaining $L{\rm{-}}1$ decoding paths, we can obtain $\mathcal{U}_{N,m}$ by selecting the $|\mathcal{U}_{N,m}|$ most reliable ones.

Further, to reduce the  searching latency and the required memory space, \cite{CRC_design} proposes a multi-level SCLD-based method by dividing $\mathcal{U}_{N,m}$ into $\left| {\mathcal{A}_m} \right|$ subsets to search, where $\mathcal{A}_m$ satisfies
\begin{equation}
\mathcal{A}_m=\{ {i {{\in}} {{\mathcal{A}}}| {{w}( \mathbf{g}_N^{(i)} ) = {d_m}} } \}
\end{equation}
where $\mathbf{g}_N^{(i)}$ is the $i$-th row vector of $\mathbf{G}_N$. The division of $\mathcal{U}_{N,m}$ is expressed as
\begin{equation}\label{divideUm}
\mathcal{U}_{N,m}{\rm{=}}\bigcup\nolimits_{i \in\mathcal{A}_m} {\mathcal{U}_{N,m}^{(i) }}
\end{equation}
where $\mathcal{U}_{N,m}^{(i) }$ is the set of vector $u_1^N$ that satisfies the following two conditions:
\begin{enumerate}
  \item $u_1^{i-1}{\rm{=}}\mathbf{0}_{i-1}$, $u_i{\rm{=}}1$, $\mathbf{u}_{\mathcal{A}_i}{\rm{\in}}\mathcal{B}^{|{\mathcal{A}_i}|}$ and $\mathbf{u}_{\mathcal{A}^c_i}{\rm{=}}\mathbf{0}_{|{\mathcal{A}^c_i}|}$, with $\mathcal{A}_{i}{\rm{=}}\mathcal{A}{\rm{\cap}}\left[\kern-0.15em\left[ {i{\rm{+}}1,N}
 \right]\kern-0.15em\right]$ and $\mathcal{A}^c_{i}{\rm{=}}\mathcal{A}^c{\rm{\cap}}\left[\kern-0.15em\left[ {i{\rm{+}}1,N}
 \right]\kern-0.15em\right]$.
  \item $w(u_1^N\mathbf{G}_N){\rm{=}}w(\mathbf{g}_N^{(i)})$.
\end{enumerate}
Based on the one to one relationship between the codeword with its information vector, when $\mathcal{U}_{N,m}$ is searched, the set of polar codewords with MHW is also derived.
\section{Features of PC-MHW as Output of Multi-Level SCLD}
\subsection{High Level Description of Enumeration}
In the following, we shall use the expression of \emph{searched path} to refer to the decoded path of SCLD under $y_1^N{\rm{=}}\mathbf{1}_N$. Now that the information vectors in $\mathcal{U}_{N,m}^{(i)}$ can be searched by the multi-level SCLD \cite{CRC_design}, they should have the same feature that can be captured by the decoder. Actually, \cite{CRC_design} has proved that the PMs of all the searched paths in $\mathcal{U}_{N,m}^{(i)}$ are equal. In this paper, we further prove that for any searched path in $\mathcal{U}_{N,m}^{(i)}$, its RDS is fixed to $\{i\}$ (Theorem 7). In general AWGN channel, where none of the decoding LLR is 0-valued, there can exist only one decoding path of SCLD whose RDS is $\{i\}$. However, under $y_1^N{\rm{=}}\mathbf{1}_N$, there can be $|\mathcal{U}_{N,m}^{(i)}|{\rm{\geq}}1$ such paths. This implies that for any searched path in $\mathcal{U}_{N,m}^{(i)}$, some decoding LLRs would be 0. In other words, the number of 0-valued decoding LLR should be associated with $|\mathcal{U}_{N,m}^{(i)}|$.

Guided by this conclusion, in Section IV we further prove that for all the searched paths in $\mathcal{U}_{N,m}^{(i)}$, the location sets of their 0-valued decoding LLR are identical (Theorem 16). Such set, denoted by $\mathcal{I}_i^0$, can be determined based on a concept of \emph{zero-capacity bit-channel associated with $i$}. Using $\mathcal{I}_i^0$, we can give the upper bound of $|\mathcal{U}_{N,m}^{(i)}|$ (Theorem 17). Since $\mathcal{U}_{N,m}{\rm{=}}\bigcup\nolimits_{i \in\mathcal{A}_m} {\mathcal{U}_{N,m}^{(i) }}$, by adding up the upper bound of each subset, the upper bound of $|\mathcal{U}_{N,m}|$ follows.

Recalling that  $\mathcal{U}_{N,m}$ can be divided into several subsets according to index set $\mathcal{A}_m$, in Section V, we adopt this idea and further divide $\mathcal{U}_{N,m}^{(i)}$ into several small subsets according to index set $\mathcal{I}_i^0$ [see equation (\ref{divideUmi})]. Based on this division, we propose a searching strategy for PC-MHW whose required list size is less than half of that required in the existing methods (Algorithm 1).

The theoretical analysis of this paper is based on three steps of simplification. First, by introducing a retracing SCLD, we can concentrate on one single searched path, but instead of all of them output by SCLD. Then, we divide the searching process of one single path into two phases and prove that the 0-valued decoding LLR can only be generated at the second phase. Thus, we only need to focus on the second phase when seeking the locations of 0-valued decoding LLRs in a searched path. Finally, the third simplification is the decomposition of the second searching phase. Note that this decomposition,  based on the recursive structure of polar codes, is not a new idea: it was applied in \cite{SCcodetree} to simplify the process of SCD.
\subsection{Retracing SCLD Depicted on Code Tree}
For any decoding path $\hat u_N[l]_{1}^N$, its message updating rules in SCLD are identical with those used in SCD except for the final hard bit decision rule. Actually, if $\overline{\mathcal{A}}_N[l]{\rm{\cap}}\mathcal{A}{\rm{=}}\phi$, $\hat u_N[l]_{1}^N$ can also be seen as the output of SCD. Based on this, we propose a \emph{retracing SCLD} (RSCLD) which performs the same message updating operation as SCD but with a determined decoding output.

We will use the code tree to describe the retracing process. In this way, the retracing process is actually a process of updating the two kinds of message vectors for each node.  The updating order is identical with SCD.  To simplify notation, in the RSCLD we still use ${{\alpha[V_\lambda^{(j)}]} }$ and ${{\beta[V_\lambda^{(j)}]} }$ to respectively denote the LLR and codeword vector of node $V_\lambda^{(j)}$ like in SCD.

RSCLD outputs only one single path and its only difference from SCD is the process of hard bit decision. Using RSCLD, we just retrace the generating process of a known decoded path of SCLD. Thus, $\overline{\mathcal{A}}_N[l]$ is assumed to be known. Concretely, on the code tree, the RSCLD starts with setting ${{\alpha[V_n^{(1)}]_i} }{\rm{=}}{{2{y_i}} \mathord{\left/
 {\vphantom {{2{y_i}} {{\sigma ^2}}}} \right.
 \kern-\nulldelimiterspace} {{\sigma ^2}}}$, $i{\rm{\in}}\left[\kern-0.15em\left[ {1,N}
 \right]\kern-0.15em\right]$. Then, it performs the same decoding procedure as SCD except for the leaf nodes. At stage 0, it should adopt the following decision rule
 \begin{equation}\label{HBDSCL}
\hat u_N[l]_j {\rm{=}}\beta[V_0^{(j)}]{\rm{=}}\left\{ \begin{gathered}
  1{\rm{-}}h({ {\alpha[V_0^{(j)}]} }), \; j \in \overline{\mathcal{A}}_N[l]\cap\mathcal{A} \hfill \\
  h({ {\alpha[V_0^{(j)}]} }),\quad\; \;\operatorname{others}  \hfill \\
\end{gathered}  \right.
\end{equation}

\emph{Definition 3 (GAN): On the code tree, for any non-root node  $V_\lambda^{(j)}$, there exists one edge that can connect $V_\lambda^{(j)}$ with the root node. The generalized ancestor nodes (GAN) of $V_\lambda^{(j)}$ are the nodes on such edges (including $V_\lambda^{(j)}$ itself). The set of the GAN of  $V_\lambda^{(j)}$ is denoted by $\mathcal{G}[V_\lambda^{(j)}]$.}

\emph{Example 2: On the 8-length code tree, shown in Fig.1(a), we use the filled cycles to signify the nodes in $\mathcal{G}[V_0^{(3)}]$. At any stage larger than $\lambda$, there is one and only one GAN of $V_\lambda^{(j)}$.}

We can conclude some properties about GAN:\\
\emph{Property 1:} For any node $V_\lambda^{(j)}$, $\lambda{\rm{\in}}\left[\kern-0.15em\left[ {0,n{\rm{-}}1}
 \right]\kern-0.15em\right]$, $j{\rm{\in}}\left[\kern-0.25em\left[ {1,2^{N-\lambda}}
 \right]\kern-0.25em\right]$, to determine $\alpha[{V_\lambda^{(j)}}]$, all the LLR vectors of its GAN should be calculated in advance.\\
 \emph{Property 2:} Given a leaf node $V_0^{(i)}$, $i{\rm{\in}}\left[\kern-0.15em\left[ {1,N}
 \right]\kern-0.15em\right]$, if it participates in updating $\beta[V_\lambda^{(j)}]$, $V_\lambda^{(j)}{\rm{\in}}\mathcal{G}[V_0^{(i)}]$. \\
 \emph{Property 3:} A node is the GAN of both its two child nodes. \\
\emph{Property 4:} Given 3 nodes $V_{\lambda_1}^{({j_1})}$, $V_{\lambda_2}^{({j_2})}$ and $V_{\lambda_3}^{({j_3})}$, $\lambda_1{\rm{>}}\lambda_2{\rm{>}}\lambda_3$, if $V_{\lambda_1}^{({j_1})}{\rm{\in}}\mathcal{G}[V_{\lambda_2}^{({j_2})}]$ and $V_{\lambda_2}^{({j_2})}{\rm{\in}}\mathcal{G}[V_{\lambda_3}^{({j_3})}]$, then $V_{\lambda_1}^{({j_1})}{\rm{\in}}\mathcal{G}[V_{\lambda_3}^{({j_3})}]$. \\
\subsection{Division of Searching Process}
From \cite{CRC_design}, we can find that $\mathcal{U}_{N,m}$ can be divided by (\ref{divideUm}) to search. This implies that the information vectors included in $\mathcal{U}_{N,m}^{(i)}$, $i{\rm{\in}}\mathcal{A}_m$, should share some common features when searched by SCLD. To explore such features, we shall first focus on a single searched path in $\mathcal{U}_{N,m}^{(i)}$.

For any searched path $\hat u_N[l]_1^N{\rm{\in}}\mathcal{U}_{N,m}^{(i) }$, its decoding process can be divided into 2 phases, i.e., the decoding of $\hat u_N[l]_1^{i}{\rm{=}}[\mathbf{0}_{i-1},1]$ (the 1st phase) and the decoding of $\hat u_N[l]_{i+1}^{N}$ (the 2nd phase). In other words, the first phase starts with feeding the SCLD by $y_1^N$ and ends up with decoding $u_i$. The remaining part of decoding process is the second phase. Note that in the following when referring to the decoding phase, we only consider in a single path.

If we use RSCLD to retrace the decoding process of any searched path $\hat u_N[l]_1^N{\rm{\in}}\mathcal{U}_{N,m}^{(i) }$, we can summarize the features of its first decoding phase in the following lemmas.

\emph{Lemma 4: For any nodes $V_\lambda^{(j)}$, if $\alpha[V_\lambda^{(j)}]$ is updated at the first decoding phase, all the LLRs in $\alpha[V_\lambda^{(j)}]$ are identical and positive, i.e., $\alpha[V_\lambda^{(j)}]_1{\rm{=}}\alpha[V_\lambda^{(j)}]_2{\rm{=}}...{\rm{=}}\alpha[V_\lambda^{(j)}]_{2^\lambda}{\rm{>}}0$. If $\beta[V_\lambda^{(j)}]$ is updated at the first decoding phase, we have $\beta[V_\lambda^{(j)}]{\rm{=}}\mathbf{0}_{2^\lambda}$}.

\emph{Proof:} This can be easily verified by the message updating rules under $\alpha[V_n^{(1)}]{\rm{=}}y_1^N{\rm{=}}\mathbf{1}_N$.$\hfill \blacksquare$

\emph{Lemma 5: For any searched path $\hat u_N[l]_1^N{\rm{\in}}\mathcal{U}_{N,m}^{(i) }$, its RDS should include $i$, i.e.,  $\overline{{\mathcal{A}}}_N[l]{\rm{\supseteq}}\{i\}$.}

\emph{Proof:} From \emph{Lemma 4}, all the decoding LLRs updated at the first decoding phase should be positive. Considering $\hat u_N[l]_1^{i}{\rm{=}}[\mathbf{0}_{i-1},1]$, the theorem follows.$\hfill \blacksquare$

Let $\mathcal{U}_N^{(i)}$ be the set of vector $u_1^N$ that satisfies $u_1^{i-1}{\rm{=}}\mathbf{0}_{i-1}$, $u_i{\rm{=}}1$, $\mathbf{u}_{\mathcal{A}_r}{\rm{\in}}\mathcal{B}^{|{\mathcal{A}_r}|}$, $\mathbf{u}_{\mathcal{A}^c_r}{\rm{=}}\mathbf{0}_{|{\mathcal{A}^c_r}|}$. Clearly, $\mathcal{U}_{N,m}^{(i)}{\rm{\subseteq}}\mathcal{U}_N^{(i)}$. When using the multi-level SCLD in \cite{CRC_design} to search $\mathcal{U}_{N,m}^{(i)}$, with $i{\rm{\in}}\mathcal{A}_m$, the work what the decoder actually do is to recognize the paths in $\mathcal{U}_{N,m}^{(i)}$ from all the paths in $\mathcal{U}_N^{(i)}$. Since all the vectors in $\mathcal{U}_N^{(i) }$ (including $\mathcal{U}_{N,m}^{(i) }$) share the same first $i$ bits, i.e., $[\mathbf{0}_{i-1},1]$, the first decoding phase of all the paths in $\mathcal{U}_N^{(i) }$ are identical. That's to say, the second decoding phase is the crux to search $\mathcal{U}_{N,m}^{(i) }$ by multi-SCLD and should be analyzed emphatically.
\subsection{Characteristics of Searched Path in $\mathcal{U}_{N,m}^{(i) }$}
In this part, we will focus on the second decoding phase to review how the PC-MHW can be searched by the multi-level SCLD.

\emph{Theorem 6: For any two different searched paths, if their corresponding codewords have the same weight, their PMs should be equal and vice versa}.

\emph{Proof:} This theorem can be easily proved by [6, lemma 3].$\hfill \blacksquare$

\emph{Theorem 7: For any searched path $\hat u_N[l]_{1}^N$ that satisfies $\hat u_N[l]_{1}^N{\rm{\in}}\mathcal{U}_{N,m}^{(i) }$,
 we have $\overline{\mathcal{A}}_N[l]=\{i\}$}.

\emph{Proof:} Please see Appendix A.$\hfill \blacksquare$

\emph{Theorem 8: For any searched path $\hat u_N[l]_1^N{\rm{\in}}\mathcal{U}_{N,m}^{(i) }$, if it has $\overline{\mathcal{A}}_N[l]{\rm{=}}\left\{ {i} \right\}$, $w(\hat u_N[l]_1^N\mathbf{G}_N){\rm{=}}w(\mathbf{g}_N^{(i)})$.}

\emph{Proof:} Since $\overline{\mathcal{A}}_N[l]{\rm{=}}\left\{ {i} \right\}$, from (\ref{PM}), the PM of $\hat u_N[l]_1^N$ is $\left|\alpha[V_0^{(i)}]\right|$. Based on \emph{Theorem 7}, the PM of the searched path $[\mathbf{0}_{i-1},1,\mathbf{0}_{N-i}]$ is also $\left|\alpha[V_0^{(i)}]\right|$. From \emph{Theorem 6}, we have
 \begin{equation}\label{theo3}
w(\hat u_N[l]_1^N\mathbf{G}_N){\rm{=}}w([\mathbf{0}_{i-1},1,\mathbf{0}_{N-i}]\mathbf{G}_N){\rm{=}}w(\mathbf{g}_N^{(i)}).
 \end{equation}

Therefore, the theorem is true. $\hfill \blacksquare$

Note that if the equation of PM is changed, e.g., using (10) in \cite{LLRSCL}, \emph{Theorem 8} still holds. This is because such change will not effect the decoding result and the RDS of path will not change.

From the above, we can draw the following conclusions

 \begin{itemize}
   \item For any searched path $\hat u_N[l]_1^N{\rm{\in}}\mathcal{U}_{N,m}^{(i)}$, its decoding bits obtained at the second decoding phase, i.e., $\hat u_N[l]_{i+1}^N$, are hard decided based on the  decoding LLR.
   \item The necessary and sufficient condition for a searched path $\hat u_N[l]_1^N$ belonging to $\mathcal{U}_{N,m}^{(i)}$, $i{\rm{\in}}\mathcal{A}_m$, is $\overline{\mathcal{A}}_N[l]{\rm{=}}\{i\}$, i.e.,
 \begin{equation}\label{conclusion}
 \hat u_N[l]_1^N{\rm{\in}}\mathcal{U}_{N,m}^{(i)}\Leftrightarrow\overline{\mathcal{A}}_N[l]{\rm{=}}\{i\}, \quad i{\rm{\in}}\mathcal{A}_m
 \end{equation}
  \item The number of $\mathcal{U}_{N,m}^{(i)}$ equals to the number of the searched paths whose RDS is $\{i\}$.
 \end{itemize}

In general AWGN channel, where the noise can not be negligible and none of the decoding LLRs is 0-valued, there can only be one decoding path of SCLD whose RDS is $\{i\}$. However, given $y_1^N{\rm{=}}\mathbf{1}_N$ and decoded trajectory $\hat u_1^{i}{\rm{=}}[\mathbf{0}_{i-1},1]$, some decoding LLRs for $u_{i+1}^N$ could be 0-valued and their corresponding decoding bits, whether decoded as 1 or 0, can be regarded to be hard decided by the suggestion of the LLR. Thus, at a position of 0-valued decoding LLR, the current decoding trajectory can be split into 2 threads with no change of the original RDS. This is essentially why there are multiple searched paths whose RDS is $\{i\}$. This implies that given $y_1^N$ and $\hat u_1^{i}{\rm{=}}[\mathbf{0}_{i-1},1]$, the number of the zero-valued decoding LLR updated in the second decoding phase may determine $\left|\mathcal{U}_{N,m}^{(i)}\right|$.
\section{Enumerator of PC-MHW}
In previous, RSCLD made it possible to simplify the analysis of the entire searching process into the analysis of a single path in $\mathcal{U}^{(i)}_{N,m}$. Since the path is chosen randomly,  the analysis can reflect the commonality of all paths in $\mathcal{U}^{(i)}_{N,m}$. We obtained that the locations of the 0-valued decoding LLRs would associate with $|\mathcal{U}_{N,m}|$. To seek such locations, in this part, we will further simplify the analysis by decomposition of the second searching phase of any path in $\mathcal{U}^{(i)}_{N,m}$.
\subsection{Decomposition of Second Searching Process}
When searching any path $\hat u_N[l]_1^N{\rm{\in}}\mathcal{U}_{N,m}^{(i)}$, $\hat u_N[l]_i{\rm{=}}1$ is the only reason to cause some subsequent decoding LLRs to be 0-valued. Hence, the 0-valued decoding LLRs can only exist in the second decoding phase. To determine their positions, we will disassemble the second decoding phase.

The  decomposition can be visualized on the code tree by dividing the nodes whose LLR vector is updated at the second phase into several subcode-trees. Since the code tree used to represent decoder is full binary, the subcode-trees obtained by decomposition can be uniquely identified by their leaf nodes. That's to say, dividing all the nodes whose LLR vector is updated at the second phase is tantamount to dividing the $N{\rm{-}}i$ leaf nodes following $V_0^{(i)}$, i.e.,  $V_0^{(i+1)}$ to $V_0^{(N)}$. We will introduce the dividing method and prove its rationality in \emph{Theorem 10}. Before that, we first give a lemma to prove the existence of the decomposition for any $i{\rm{\in}}\left[\kern-0.15em\left[ {1,2^n{\rm{-}}1}
 \right]\kern-0.15em\right]$.

\emph{Lemma 9: Given any integer $i{\rm{\in}}\left[\kern-0.15em\left[ {1,2^n{\rm{-}}1}
 \right]\kern-0.15em\right]$, we have
  \begin{equation}\label{integer}
  2^n{\rm{-}}i{\rm{=}}\gamma_i(n{\rm{-}}w(\mathbf{b}_n\left( {i {\rm{-}} 1} \right))){\rm{=}}\gamma_i(\left| {{{{\mathcal{P}}}_0}\left[ {{\mathbf{b}_n}\left( {i {\rm{-}} 1} \right)} \right]} \right|)
  \end{equation}
where
\begin{equation}\label{gamma}
\gamma_i(x){\rm{\triangleq}}\sum\nolimits_{k = 1}^{x} {{2^{{\mathcal{P}_0}{{\left[ {{\mathbf{b}_n}\left( {i - 1} \right)} \right]}_k} - 1}}}.
  \end{equation}}

\emph{Proof:} Since  $2^n{\rm{-}}i{\rm{=}}(2^n{\rm{-}}1){\rm{-}}(i{\rm{-}}1)$, this lemma can be easily proved by expanding $2^n{\rm{-}}1$ and  $i{\rm{-}}1$ binary.$\hfill \blacksquare$

\emph{Lemma 9} implies that any positive integer $2^n{\rm{-}}i$ can be broken into
 $\left| {{{{\mathcal{P}}}_0}\left[ {{\mathbf{b}_n}\left( {i {\rm{-}} 1} \right)} \right]} \right|$ smaller positive integers. Thus, for any leaf node with index $i{\rm{\in}}\left[\kern-0.15em\left[ {1,N {\rm{-}} 1}
 \right]\kern-0.15em\right]$ (i.e., $V_0^{(i)}$), we can divide its subsequent $N{\rm{-}}i$ ones into  $|{{{\mathcal{P}}}_0}\left[ {{\mathbf{b}_n}\left( {i {\rm{-}} 1} \right)} \right]|$ parts. For the $k$-th part, $k{\rm{\in}}\left[\kern-0.15em\left[ {1,|{{{\mathcal{P}}}_0}\left[ {{\mathbf{b}_n}\left( {i {\rm{-}} 1} \right)}\right]| }
 \right]\kern-0.15em\right]$, let $\mathcal{I}_i(k)$ denote the index set of the leaf nodes in it. Specifically, $\mathcal{I}_i(k)$ has $2^{{{{\mathcal{P}}}_0}\left[ {{\mathbf{b}_n}\left( {i {\rm{-}} 1} \right)} \right]_k-1}$ consecutive indices and is arrayed in the ascending order
  \begin{equation}\label{subcodetreeindex}
{{{\mathcal{I}}}_i}(k){\rm{=}}\left\{ \begin{gathered}
  \left[\kern-0.15em\left[ {i {\rm{+}} 1,i {\rm{+}} \gamma_i(1)}
 \right]\kern-0.15em\right], \quad\quad\quad\quad\;\;\,\, k{\rm{=}}1 \hfill \\
  \left[\kern-0.15em\left[ {i {\rm{+}} 1 {\rm{+}}\gamma_i(k{\rm{-}}1)  ,i{\rm{+}}\gamma_i(k) }
 \right]\kern-0.15em\right],\; \;k{\rm{\geq}}2  \hfill \\
\end{gathered}  \right.
\end{equation}

\emph{Theorem 10: The nodes whose LLR vector is calculated at the second phase can be divided into $\left| {{{{\mathcal{P}}}_0}\left[ {{\mathbf{b}_n}\left( {i {\rm{-}} 1} \right)} \right]} \right|$ subcode-trees. For the $k$-th subcode-tree, $k{\rm{\in}}\left[\kern-0.15em\left[ {1,\left|{{{\mathcal{P}}}_0}\left[ {{\mathbf{b}_n}\left( {i {\rm{-}} 1} \right)} \right]\right|}
 \right]\kern-0.15em\right]$, the index set of the  leaf nodes is $\mathcal{I}_i(k)$. By doing this, we obtain that:\\
1)  The $k$-th subcode-tree is rooted at node $V_{\lambda(i,k)}^{(j(i,k))}$, where
\begin{equation*}
\lambda(i,k){\rm{=}}{{{\mathcal{P}}}_0}\left[ {{\mathbf{b}_n}\left( {i {\rm{-}} 1} \right)} \right]_k{\rm{-}}1,\; j(i,k){\rm{=}}\frac{i+\gamma_i(k)}{2^{{{{\mathcal{P}}}_0}\left[ {{\mathbf{b}_n}\left( {i {\rm{-}} 1} \right)} \right]_k-1}}.
\end{equation*}\\
2) For any node whose LLR vector is updated at the second phase, it must be involved in one and only one of the $\left| {{{{\mathcal{P}}}_0}\left[ {{\mathbf{b}_n}\left( {i {\rm{-}} 1} \right)} \right]} \right|$ subcode-trees.}

\emph{Proof:} For the first problem, from (\ref{subcodetreeindex}), we can find that the number of the leaf nodes involved in the $k$-th subcode-tree is $2^{{{{\mathcal{P}}}_0}\left[ {{\mathbf{b}_n}\left( {i {\rm{-}} 1} \right)} \right]_k{\rm{-}}1}$. Considering the subcode-tree is full binary, the depth of the subcode-tree is ${{{{\mathcal{P}}}_0}\left[ {{\mathbf{b}_n}\left( {i {\rm{-}} 1} \right)} \right]_k{\rm{-}}1}$. Thus, $\lambda(i,k){\rm{=}}{{{\mathcal{P}}}_0}\left[ {{\mathbf{b}_n}\left( {i {\rm{-}} 1} \right)} \right]_k{\rm{-}}1$. Similarly, since $i{\rm{+}}\gamma_i(k)$ is the maximum index in $\mathcal{I}_i(k)$, we have  $j(i,k){\rm{=}}\frac{i+\gamma_i(k)}{2^{{{{\mathcal{P}}}_0}\left[ {{\mathbf{b}_n}\left( {i {\rm{-}} 1} \right)} \right]_k-1}}$.

Then, we will deal with the second problem. Assume that $\alpha[V_\lambda^{(j)}]$ is updated at the second phase. Clearly, node $V_\lambda^{(j)}$ should be GAN of at least one leaf node, denoted by $V_0^{(j_0)}$. Since the subcode-trees are full binary, if $V_\lambda^{(j)}$ is excluded from the $\left| {{{{\mathcal{P}}}_0}\left[ {{\mathbf{b}_n}\left( {i {\rm{-}} 1} \right)} \right]} \right|$ subcode-trees obtained by decomposition, then the leaf node $V_0^{(j_0)}$ should be also excluded. This contradicts with the fact obtained from \emph{Lemma 9}, i.e., none of the leaf nodes is excluded from the obtained subcode-trees. Thus, all the nodes whose LLR vector is updated at the second phase are involved in by the decomposition. Similarly, if $V_\lambda^{(j)}$ belongs to two different subcode-trees simultaneously, leaf node $V_0^{(j_0)}$ should also belong to two subcode-trees. This also contradicts with the fact. Thus, the theorem follows.$\hfill \blacksquare$

Note that $\lambda(i,k){\rm{=}}0$ indicates that $i$ is odd and $k{\rm{=}}1$. Conversely,  $k{\rm{=}}1$ does not necessarily lead to
$\lambda(i,k){\rm{=}}0$.

\emph{Example 3: In Fig.1(a), if $i{\rm{=}}1$, three subcode-trees obtained by the decomposition according to Theorem 10 are boxed out by the dashed line. The LLR vector of any excluded node is updated  at the first phase.
Note that if $\lambda(i,k){\rm{=}}0$, the subcode-tree is a single leaf node $V_0^{(i+1)}$.}

It is worth mentioning that this dividing is just a method to facilitate analysis and will not change or omit any process of the original searching. That's to say, it would enable some procedures and factors in the second decoding phase to be ignored in the analysis, but they  still exist in the actual decoding process. Specifically, in each local decoder (or, subcode-tree), we can deem that only the LLR vector of the root node is effected by its previous decoding while the other nodes are only effected by their root node. Therefore, once the LLR vector of the root node is determined, i.e., on activation of each local decoder, the impact of the previous decoding can be ignored in all its following message updating process in this local decoder.
\subsection{Input Vector of Subcode-Tree}
Based on the above decomposition, for any searched path in $\mathcal{U}_{N,m}^{(i)}$, determining\footnote{We aim to give a theoretical  prediction to determine the locations of zero-valued decoding LLRs, but instead of directly using the results of the decoder. Actually, in this section we will prove that for any searched path in $\mathcal{U}_m^{(i)}$, the locations of 0-valued decoding LLRs are fixed.} its locations of the 0-valued decoding LLRs from the whole $N$ locations can be simplified by first determining them in any one of the $\left| {{{{\mathcal{P}}}_0}\left[ {{\mathbf{b}_n}\left( {i {\rm{-}} 1} \right)} \right]} \right|$ local decoders. Then, in the rest $\left| {{{{\mathcal{P}}}_0}\left[ {{\mathbf{b}_n}\left( {i {\rm{-}} 1} \right)} \right]} \right|{\rm{-}}1$ ones, we can obtain them similarly. The crux of such simplification is to determine the input vector for each local decoders so that they can be regarded to be independent. We can use RSCLD to retrace the decoding process and draw some conclusions about the root node of the local decoder, which are given as follows.

\emph{Theorem 11: For any $i{\rm{\in}}\left[\kern-0.15em\left[ {1,N {\rm{-}} 1}
 \right]\kern-0.15em\right]$ and $k{\rm{\in}}\left[\kern-0.15em\left[ {1,|{{{\mathcal{P}}}_0}\left[ {{\mathbf{b}_n}\left( {i {\rm{-}} 1} \right)}\right]| }
 \right]\kern-0.15em\right]$, $j(i,k)$ is even.}

 \emph{Proof:} The binary expansion of $i{\rm{+}}\gamma_i(k)$ is $\mathbf{b}_n(i+\gamma_i(k)){\rm{=}}[b_{i{\rm{+}}\gamma_i(k),1},b_{i+\gamma_i(k),2},...,b_{i+\gamma_i(k),n}]$.

Meanwhile, we have
 \begin{equation}\label{Theorem10}
 \mathbf{b}_n(i{\rm{+}}\gamma_i(k))= \mathbf{b}_n(i{\rm{-}}1)+\mathbf{b}_n(\gamma_i(k))+\mathbf{b}_n(1)
  \end{equation}
Substituting (\ref{gamma}) into (\ref{Theorem10}), we have $[b_{i+\gamma_i(k),1},b_{i+\gamma_i(k),2},...,b_{i+\gamma_i(k),{{{{\mathcal{P}}}_0}\left[ {{\mathbf{b}_n}\left( {i {\rm{-}} 1} \right)} \right]_k}}]{\rm{=}}\mathbf{0}_{{{{{\mathcal{P}}}_0}\left[ {{\mathbf{b}_n}\left( {i {\rm{-}} 1} \right)} \right]_k}}$.
 This implies that $ \frac{i+\gamma_i(k)}{2^{{{{\mathcal{P}}}_0}\left[ {{\mathbf{b}_n}\left( {i {\rm{-}} 1} \right)} \right]_k}}{\rm{=}}\frac{j(i,k)}{2}{\rm{\in}}\mathcal{Z}^*.$
  The theorem is proved. $\hfill \blacksquare$

\emph{Theorem 11} indicates that when using RSCLD to retrace the searching process of any path $\hat u_N[l]_1^N{\rm{\in}}\mathcal{U}_m^{(i)}$, $\alpha[V_{\lambda(i,k)}^{(j(i,k))}]$ can only be calculated by  (\ref{LLRbit}) in which $\alpha[V_{\lambda(i,k)+1}^{({{j\left( {i,k} \right)} \mathord{\left/
 {\vphantom {{j\left( {i,k} \right)} 2}} \right.
 \kern-\nulldelimiterspace} 2})}]$ and $\beta[V_{\lambda(i,k)}^{(j(i,k)-1)}]$ will participate. By recalling the updating rule of $\beta[V_{\lambda(i,k)}^{(j(i,k)-1)}]$, i.e., equation (\ref{debit}), we can easily obtain that
 \begin{equation}\label{rootbeta}
\beta[V_{\lambda(i,k)}^{(j(i,k)-1)}] {\rm{=}} \left\{ \begin{gathered}
 1,\quad  \quad\quad\quad\quad\quad\quad\quad\quad\quad\quad\quad\lambda(i,k){\rm{=}}0 \hfill \\
  \hat u_N[l]_{\min({{{\mathcal{I}}}_i}(k))-2^{\lambda(i,k)}}^{\min({{{\mathcal{I}}}_i}(k))-1}\mathbf{G}_{2^{\lambda(i,k)}},\quad\,\,\,\lambda(i,k){\rm{>}}0\; \hfill\\
\end{gathered}  \right.
\end{equation}

It can be verified that
 \begin{equation*}
\min({{{\mathcal{I}}}_i}(k))-2^{\lambda(i,k)}\leq i\leq\min({{{\mathcal{I}}}_i}(k))-1.
 \end{equation*}
Thus, $u_N[l]_i$ will participate in the calculation of $\beta[V_{\lambda(i,k)}^{(j(i,k)-1)}]$. From \emph{Property  2}, $V_{\lambda(i,k)}^{(j(i,k)-1)}{\rm{\in}}\mathcal{G}(V_{0}^{(i)})$.

As the parent node of  $V_{\lambda(i,k)}^{(j(i,k)-1)}$,  $V_{\lambda(i,k)+1}^{({{j\left( {i,k} \right)} \mathord{\left/
 {\vphantom {{j\left( {i,k} \right)} 2}} \right.
 \kern-\nulldelimiterspace} 2})}$ also should be the GAN of $V_{0}^{(i)}$, and thus $\alpha[V_{\lambda(i,k)+1}^{({{j\left( {i,k} \right)} \mathord{\left/
 {\vphantom {{j\left( {i,k} \right)} 2}} \right.
 \kern-\nulldelimiterspace} 2})}]$ is updated at the first decoding phase. From \emph{Lemma 4}, we can obtain that
\begin{equation}\label{rootLLR}
\alpha[V_{\lambda(i,k)+1}^{({{j\left( {i,k} \right)} \mathord{\left/
 {\vphantom {{j\left( {i,k} \right)} 2}} \right.
 \kern-\nulldelimiterspace} 2})}]=x\cdot\mathbf{1}_{2^{\lambda{(i,k)+1}}}
\end{equation}

Substituting (\ref{rootbeta}) and (\ref{rootLLR}) into (\ref{LLRbit}), we can conclude that
\begin{equation}
\begin{gathered}\label{LLRlocaldecoder}
\begin{aligned}
\alpha[V_{\lambda(i,k)}^{(j(i,k))}]
{\rm{=}}\left\{ \begin{gathered}
  2x{\rm{\cdot}}(\mathbf{1}_{2^{\lambda{(i,k)}}}{\rm{-}}\hat u_N[l]_{\min({{{\mathcal{I}}}_i}(k))-2^{\lambda(i,k)}}^{\min({{{\mathcal{I}}}_i}(k))-1}\mathbf{G}_{2^{\lambda(i,k)}}), \;\;\quad\; \lambda(i,k){\rm{>}}0 \hfill \\
  0,\,\;\, \;\quad\quad\quad\quad\quad\quad\quad\quad\quad\quad\quad\quad\quad\quad\quad\quad\quad \lambda(i,k){\rm{=}}0\; \hfill\\
\end{gathered}  \right.
\end{aligned}
\end{gathered}
\end{equation}
We can further determine the value of constant $x$. For the two updating formulae associated with LLR, i.e., (\ref{LLRnobit}) and (\ref{LLRbit}),  at the first decoding phase only  (\ref{LLRbit}) can change (or precisely, double) the LLR value of the input LLR vector. We have
\begin{equation}
\begin{gathered}
\begin{aligned}
x
{\rm{=}}\left\{ \begin{gathered}
  2^{w(\mathbf{b}_n(i{\rm{-}}1)_{\lambda(i,k)+2}^n)}, \;\;\quad\; \lambda(i,k)+1<n \hfill \\
  1,\,\;\,\, \;\quad\quad\quad\quad\quad\quad\quad \lambda(i,k)+1=n\; \hfill\\
\end{gathered}  \right.
\end{aligned}
\end{gathered}
\end{equation}

where $w(\mathbf{b}_n(i{\rm{-}}1)_{\lambda(i,k)+2}^n)$ is equal to the times of using (\ref{LLRbit}) to  obtain $\alpha[V_{\lambda(i,k)+1}^{({{j\left( {i,k} \right)} \mathord{\left/
 {\vphantom {{j\left( {i,k} \right)} 2}} \right.
 \kern-\nulldelimiterspace} 2})}]$, updating from root node $V_{n}^{(1)}$ through the GANs of $V_{\lambda(i,k)+1}^{({{j\left( {i,k} \right)} \mathord{\left/
 {\vphantom {{j\left( {i,k} \right)} 2}} \right.
 \kern-\nulldelimiterspace} 2})}$.
\subsection{Zero-Capacity Bit-Channel}
Let $\mathbf{g}_N^{(i)}(x)$ denote the vector of the first $2^x$ elements in $\mathbf{g}_N^{(i)}$. Recalling the input LLR vector of the decomposed local decoder that has been determined in (\ref{LLRlocaldecoder}), the local decoders can be seen to be independent with each other. We can first focus on any one of the local decoders. For any $i{\rm{\in}}\left[\kern-0.15em\left[ {1,N {\rm{-}} 1}
 \right]\kern-0.15em\right]$, we will define the zero-capacity bit-channels (ZCBC) as follows

\emph{Definition 12: For any $k{\rm{\in}}\left[\kern-0.15em\left[ {1,\left|{{{\mathcal{P}}}_0}\left[ {{\mathbf{b}_n}\left( {i {\rm{-}} 1} \right)} \right]\right|}
 \right]\kern-0.15em\right]$, if we use ${{{\mathcal{I}}}^0_i}(k)$ to denote the set of the $h$-th element in ${{{\mathcal{I}}}_i}(k)$, where
\begin{equation}
h\in \mathcal{P}_1[\mathbf{g}^{(i)}_N({{{{\mathcal{P}}}_0}\left[ {{\mathbf{b}_n}\left( {i {\rm{-}} 1} \right)} \right]_k{\rm{-}}1})],
\end{equation}
then the bit-channels with indices in ${{{\mathcal{I}}}^0_i}(k)$ are defined as zero-capacity bit-channels associated with $i$ (ZCBC-i).}

Note that all the $\left|{{{\mathcal{P}}}_0}\left[ {{\mathbf{b}_n}\left( {i {\rm{-}} 1} \right)} \right]\right|$ local decoders have their own respective ZCBC. Thus, the set of index of ZCBC-i, denoted by $\mathcal{I}_i^0$, can be expressed as
\begin{equation}
\mathcal{I}_i^0=\bigcup\nolimits_{k = 1}^{\left| {{{{\mathcal{P}}}_0}\left[ {{b_n}\left( {i - 1} \right)} \right]} \right|} {{{\mathcal{I}}}_i^0\left( k \right)}
\end{equation}

\emph{Example 4: For 8-length polar codes (i.e., $n{\rm{=}}3$), if $i{\rm{=}}2$, we have ${{{\mathcal{P}}}_0}\left[ {{\mathbf{b}_n}\left( {i {\rm{-}} 1} \right)} \right]{\rm{=}}\{2,3\}$. Thus, the following 6 bit-channels can be divided into $|{{{\mathcal{P}}}_0}\left[ {{\mathbf{b}_n}\left( {i {\rm{-}} 1} \right)} \right]|{\rm{=}}2$ parts.
Concretely, $\mathcal{I}_i{(1)}{\rm{=}}\{3,4\}$ and $\mathcal{I}_i{(2)}{\rm{=}}\{5,6,7,8\}$. Since
$\mathbf{g}_N^{(i)}({{{\mathcal{P}}}_0}\left[ {{\mathbf{b}_n}\left( {i {\rm{-}} 1} \right)} \right]_1{\rm{-}}1)=\mathbf{g}_N^{(i)}(2-1)=[1,1]$
 and $\mathcal{P}_1[\mathbf{g}_N^{(i)}(1)\,]=\{1,2\}$,
both the two bit-channels in $\mathcal{I}_i{(1)}$ are ZCBC-i. Similarly, since $\mathbf{g}_N^{(i)}[{{{\mathcal{P}}}_0}\left[ {{\mathbf{b}_n}\left( {i -1} \right)} \right]_2-1]=\mathbf{g}_N^{(i)}(3-1)=[1,1,0,0]$ and $\mathcal{P}_1[\mathbf{g}_N^{(i)}(2)]=\{1,2\}$, then the first two bit-channels in $\mathcal{I}_i{(2)}$ are ZCBC-i. Thus,  $\mathcal{I}_i^0=\mathcal{I}_i^0(1)\cup\mathcal{I}_i^0(2)=\{3,4\}\cup\{5,6\}=\{3,4,5,6\}$.}
\subsection{Upper Bound for Number of PC-MHW}
For bit-channel with index in $\mathcal{I}_i^0$, we call it  \emph{zero-capacity} because we will prove that for any searched path  $\hat u_N[l]_1^N{\rm{\in}}\mathcal{U}_{N,m}^{(i)}$, the set of the locations of the 0-valued decoding LLRs is $\mathcal{I}_i^0$. Before that, we first give three lemmas which will be incurred in the following discussion.

\emph{Lemma 13: Given decoded path of SCD $\hat u_1^N{\rm{\in}}\mathcal{U}_{N,m}^{(i)}$, $i{\rm{\in}}\left[\kern-0.15em\left[ {1,N}
 \right]\kern-0.15em\right]$, on the code tree, for any  $\lambda{\rm{\in}}\left[\kern-0.15em\left[ {0,n{\rm{-}}1}
 \right]\kern-0.15em\right]$ and $j_\lambda{\rm{\in}}\left[\kern-0.25em\left[ {1,2^{n-\lambda}}
 \right]\kern-0.25em\right]$, if $V_\lambda^{(j_\lambda)}{\rm{\in}}\mathcal{G}[V_0^{(j)}]$, then we have
\begin{equation}\label{lemma13}
w(\beta[V_{\lambda}^{(j_\lambda) }]){\rm{=}}{w(\mathbf{g}_N^{(i)})}/{2^{\left| {{\mathcal{P}_1}\left[ {{\mathbf{b}_n}\left( {i - 1} \right)_{\lambda  + 1}^n} \right]} \right|}}
\end{equation}}

\emph{Proof:} When $\lambda{\rm{=}}0$, we have
\begin{equation}
w(\beta[V_{\lambda}^{(j_\lambda) }]){\rm{=}}w(\beta[V_{0}^{(i) }]){\rm{=}}w([\hat u_i]){\rm{=}}1{\rm{=}}w(\mathbf{g}_N^{(i)})/2^{|\mathcal{P}_1[\mathbf{b}_n(i-1)]|}
 \end{equation}
The lemma is true. If $\lambda{\rm{=}}1$, it can be easily verified that
 \begin{equation*}
w(\beta[V_{1}^{(j_1) }]) {\rm{=}} \left\{ \begin{gathered}
 2{\rm{=}}w(\mathbf{g}_N^{(i)})/2^{{|\mathcal{P}_1[\mathbf{b}_n(i-1)]|}-1},\quad{\rm{if}}\,i\,{\rm{is}}\,{\rm{even}}   \hfill \\
 1{\rm{=}}w(\mathbf{g}_N^{(i)})/2^{{|\mathcal{P}_1[\mathbf{b}_n(i-1)]|}},\quad\quad{\rm{if}}\,i\,{\rm{is}}\,{\rm{odd}}\; \hfill\\
\end{gathered}  \right.
\end{equation*}
 Thus, it is also true for $\lambda{\rm{=}}1$. Due to the recursive structure, when $\lambda{\rm{=}}2$, we can regard the $N/2$ code-tree nodes at the stage 1 as the leaf nodes and ignore the nodes at stage 0. By doing this, the weight of the new leaf nodes is 2 (if its index is even) or 1 (if its index is odd). We can verify that this lemma is still true at stage 2. By an induction, the lemma follows. $\hfill \blacksquare$

\emph{Lemma 14:  For any $\lambda{\rm{\in}}\left[\kern-0.15em\left[ {0,n{\rm{-}}1}
 \right]\kern-0.15em\right]$,
 \begin{equation}\label{lemma14}
 w(\mathbf{g}_N^{(i)}(\lambda)){\rm{=}}w(\mathbf{g}_N^{(i)})/{2^{\left| {{\mathcal{P}_1}\left[ {{\mathbf{b}_n}\left( {i - 1} \right)_{\lambda  + 1}^n} \right]} \right|}}
  \end{equation}}

\emph{Proof:} If we regard this lemma as a special case of \emph{Lemma 13}, i.e., the decoded path $\hat u_1^N$ is $[\mathbf{0}_{i-1},1,\mathbf{0}_{N-i}]$, it follows.$\hfill \blacksquare$

\emph{Lemma 15: Under SCD, if the input vector $y_1^N$ satisfies
\begin{equation}\label{Theo13}
y_1^N=a\cdot(\mathbf{1}_N-u_1^N\mathbf{G}_N)
\end{equation}
where $u_1^N$ is any vector in $\mathcal{U}_m^{(i)}$, with $i{\rm{\in}}\left[\kern-0.15em\left[ {1,N{\rm{-}}1} \right]\kern-0.15em\right]$, and $a$ is any positive constant, then the location set of the 0-valued decoding LLR is $\mathcal{P}_1[\mathbf{g}_N^{(i)}]$.}

\emph{Proof:} Please see Appendix B.$\hfill \blacksquare$

\emph{Theorem 16: For any searched path  $\hat u_N[l]_1^N{\rm{\in}}\mathcal{U}_{N,m}^{(i)}$, the set of the locations of the 0-valued decoding LLRs is $\mathcal{I}_i^0$.}

 \emph{Proof:} Since we only consider one single path, we can use RSCLD to analyze it generating process. The 0-valued decoding LLRs can only be generated at the second decoding phase which can be divided into $\left| {{{{\mathcal{P}}}_0}\left[ {{\mathbf{b}_n}\left( {i {\rm{-}} 1} \right)} \right]} \right|$ local decoders according to \emph{Theorem 10}.

Without loss of generality, we first focus on the $k$-th local decoder, $k{\rm{\in}}\left[\kern-0.15em\left[ {1,|{{{\mathcal{P}}}_0}\left[ {{\mathbf{b}_n}\left( {i {\rm{-}} 1} \right)}\right]| }
 \right]\kern-0.15em\right]$. From previous analysis, the bits decoded by the $k$-th local decoder, i.e., the bits with indices in $\mathcal{I}_i(k)$ of path ${\hat u}_N[l]_1^N$, denoted as $\mathbf{\hat u}_N[l]_{\mathcal{I}_i(k)}$, are made hard decision according to their decoding LLRs. Thus, for $\mathbf{\hat u}_N[l]_{\mathcal{I}_i(k)}$, the $k$-th local decoder can be seen as an SCD with input  $\alpha[V_{\lambda(i,k)}^{(j\left( {i,k} \right))}]$\footnote{It is worth noting that SCD can only reserve one path, however, all the paths whose RDS is none can be regarded as the valid output of it. When none of the decoding LLR is 0, there is only one  valid output.}.

If $\lambda(i,k){\rm{>}}0$, $\hat u_N[l]_i$ is the $(i{\rm{-}}\min({{{\mathcal{I}}}_i}(k)){\rm{+}}2^{\lambda(i,k)}{\rm{+}}1)$-th bit in $\hat u_N[l]_{\min({{{\mathcal{I}}}_i}(k))-2^{\lambda(i,k)}}^{\min({{{\mathcal{I}}}_i}(k))-1}$. Meanwhile, the previous analysis has proved that $V_{\lambda(i,k)}^{(j(i,k)-1)}{\rm{\in}}\mathcal{G}(V_{0}^{(i)})$. Based on this two facts, we have
\begin{equation}
\begin{gathered}\label{theorem15prof}
\begin{aligned}
w(\mathbf{g}_{2^{\lambda(i,k)}}^{(i{\rm{-}}\min({{{\mathcal{I}}}_i}(k)){\rm{+}}2^{\lambda(i,k)}{\rm{+}}1)})
&\mathop  = \limits^{\left( a \right)}w(\mathbf{g}_{N}^{(i)}({\lambda(i,k)}))
\mathop  = \limits^{\left( b \right)} w(\mathbf{g}_N^{(i)})/2^{|\mathcal{P}_1[\mathbf{b}_n(i-1)^n_{\lambda(i,k)+1}]|}\\
&\mathop  = \limits^{\left( c \right)} w(\beta[V_{\lambda(i,k)}^{(j(i,k)-1)}])
\mathop  = \limits^{\left( d \right)}w(\hat u_N[l]_{\min({{{\mathcal{I}}}_i}(k))-2^{\lambda(i,k)}}^{\min({{{\mathcal{I}}}_i}(k))-1}\mathbf{G}_{2^{\lambda(i,k)}})
\end{aligned}
\end{gathered}
\end{equation}
where the step (a) is based on the structure of $\mathbf{G}_N$, step (b) comes from \emph{Lemma 14} based on the fact of , step (c) is obtained by \emph{Lemma 13} and step (d) is based on equation (\ref{rootbeta}).

Moreover, since $\hat u_N[l]_1^N{\rm{\in}}\mathcal{U}_{N,m}^{(i)}$, it follows that
\begin{equation}\label{theorem15prof1}
\hat u_N[l]_{\min{(\mathcal{I}_i(k))}-2^{\lambda(i,k)}}^{i-1}{\rm{=}}\mathbf{0}_{i{\rm{-}}\min({{{\mathcal{I}}}_i}(k)){\rm{+}}2^{\lambda(i,k)}}
\end{equation}
Combining (\ref{theorem15prof}) and (\ref{theorem15prof1}), we have $\hat u_N[l]_{\min({{{\mathcal{I}}}_i}(k))-2^{\lambda(i,k)}}^{\min({{{\mathcal{I}}}_i}(k))-1} \in\mathcal{U}_{2^{\lambda(i,k)},m}^{(i{\rm{-}}\min({{{\mathcal{I}}}_i}(k)){\rm{+}}2^{\lambda(i,k)}{\rm{+}}1)}$.  Recalling \emph{Lemma 15}, the location set of zero-valued decoding LLRs under the $k$-th local decoder is
\begin{equation}
\mathcal{P}_1[\mathbf{g}_{2^{\lambda(i,k)}}^{(i{\rm{-}}\min({{{\mathcal{I}}}_i}(k)){\rm{+}}2^{\lambda(i,k)}{\rm{+}}1)}]
=\mathcal{P}_1[\mathbf{g}_{N}^{(i)}({\lambda(i,k)})]
=\mathcal{P}_1[\mathbf{g}^{(i)}_N({{{{\mathcal{P}}}_0}\left[ {{\mathbf{b}_n}\left( {i {\rm{-}} 1} \right)} \right]_k{\rm{-}}1})]
\end{equation}
Based on the definition of ZCBC-i, the theorem is true in the $k$-th local decoder.

\underline{}In the case of $\lambda(i,k){\rm{=}}0$, we have $k{\rm{=}}1$. It follows that $j(i,k){\rm{=}}i{\rm{+}}1$. Meanwhile, in this case, it can be easily obtained that $\mathcal{I}_i^0(k){\rm{=}}\{i{\rm{+}}1\}$.  From (\ref{LLRlocaldecoder}) we can directly derive: $\alpha[V_{0}^{(i+1)}]{\rm{=}}0$. Thus, this theorem is also true in the case of $\lambda(i,k){\rm{=}}0$.

In any other local decoder, this conclusion can be derived by the same method. Overall,  the location set of  0-valued decoding LLRs generated at the whole decoding process is $\mathcal{I}_i^0$. The theorem follows.$\hfill \blacksquare$

\emph{Theorem 17: For polar codes with generator matrix $\mathbf{G}_N^{\mathcal{A}}$, the number of PC-MHW  is upper bounded by $\sum\nolimits_{i \in {{{\mathcal{A}}}_m}} {{2^{|{{\mathcal{I}}}_i^{0} \cap {{\mathcal{A}}}|}}}$, i.e., $|\mathcal{U}_{N,m}|{\rm{\leq}}\sum\nolimits_{i \in {{{\mathcal{A}}}_m}} {{2^{|{{\mathcal{I}}}_i^{0} \cap {{\mathcal{A}}}|}}}$.}

\emph{Proof:} We can first consider the upper bound of $|{\mathcal{U}_{N,m}^{(i) }}|$, $i{\rm{\in}}\mathcal{A}_m$. When searching ${\mathcal{U}_{N,m}^{(i) }}$, at any location  $j{\rm{\in}}\left[\kern-0.15em\left[ {1,i{\rm{-}}1} \right]\kern-0.15em\right]$, we call a decoding trajectory $\hat u_j[l]_1^j$ is \emph{valid} if its RSD is $\phi$. Meanwhile, at any $j{\rm{\in}}\left[\kern-0.15em\left[ {i,N} \right]\kern-0.15em\right]$, a  trajectory $\hat u_j[l]_1^j$ is \emph{valid} if its RSD is $\{i\}$. Based on \emph{Theorem 7-8}, when all the $N$ bits are decoded by SCLD, the set of valid trajectory is just ${\mathcal{U}_{N,m}^{(i) }}$. Therefore, this theorem can be proved by focusing on how the number of valid trajectories changes at each location.

For any location $j{\rm{\in}}\left[\kern-0.15em\left[ {1,i{\rm{-}}1}
 \right]\kern-0.15em\right]$, only one decoding trajectory, i.e., $\mathbf{0}_j$, is valid. Moreover, at location $i$, the valid trajectory is just [$\mathbf{0}_{i-1}$,1]. When $j{\rm{\in}}\left[\kern-0.15em\left[ {i{\rm{+}}1,N}
 \right]\kern-0.15em\right]$, for any input valid trajectory, e.g., $\hat u_{j-1}[l]_1^{j-1}$, we can consider the following 4 cases£º
 \begin{enumerate}
   \item If $j{\rm{\in}}\left[\kern-0.15em\left[ {i{\rm{+}}1,N}
 \right]\kern-0.15em\right]{\rm{\cap}}\mathcal{A}$ and the decoding LLR of $u_j$ is zero, then $\hat u_{j-1}[l]_1^{j-1}$ will be split into two valid trajectories, i.e., $[\hat u_{j-1}[l]_1^{j-1},0]$ and $[\hat u_{j-1}[l]_1^{j-1},1]$.
   \item If $j{\rm{\in}}\left[\kern-0.15em\left[ {i{\rm{+}}1,N}
 \right]\kern-0.15em\right]{\rm{\cap}}\mathcal{A}$ and the decoding LLR of $u_j$ is non-zero, then $\hat u_{j-1}[l]_1^{j-1}$ can educe only one valid path.
   \item If $j{\rm{\in}}\left[\kern-0.15em\left[ {i{\rm{+}}1,N}
 \right]\kern-0.15em\right]{\rm{\cap}}\mathcal{A}^c$ and the decoding LLR of $u_j$ is negative, then $\hat u_{j-1}[l]_1^{j-1}$ can not educe any valid path in all the subsequent decoding steps in that its RDS will become $\{i,j\}$.
   \item If $j{\rm{\in}}\left[\kern-0.15em\left[ {i{\rm{+}}1,N}
 \right]\kern-0.15em\right]{\rm{\cap}}\mathcal{A}^c$ and the decoding LLR of $u_j$ is non-negative, then $\hat u_{j-1}[l]_1^{j-1}$ will educe only one valid path, i.e., [$\hat u_{j-1}[l]_1^{j-1}$,0].
 \end{enumerate}

 In \emph{Theorem 16} we proved that the location set of zero-valued decoding LLRs of any searching path in $\mathcal{U}_{N,m}^{(i)}$ should be fixed to $\mathcal{I}_i^{0}$. Thus, for the 4 cases listed above, only in case 1), i.e., at any location in $\mathcal{I}_i^{0}{\rm{\cap}}\mathcal{A}$, the number of the valid trajectories should be doubled.
  Meanwhile, at the other locations such number will stay the same [in case 2) or 4)] or decrease [in case 3)] compared with the last location.

Therefore, we can conclude that  $|\mathcal{U}^{(i)}_{N,m}|{\rm{\leq}}  {{2^{|{{\mathcal{I}}}_i^{0} \cap {{\mathcal{A}}}|}}}$. Recalling that $\mathcal{U}_{N,m}{\rm{=}}\bigcup\nolimits_{i \in\mathcal{A}_m} {\mathcal{U}_{N,m}^{(i) }}$, we have
\begin{equation*}
|\mathcal{U}_{N,m}|=\sum\nolimits_{i \in\mathcal{A}_m} |{\mathcal{U}_{N,m}^{(i) }}|{\rm{\leq}}\sum\nolimits_{i \in {{{\mathcal{A}}}_m}} {{2^{|{{\mathcal{I}}}_i^{0} \cap {{\mathcal{A}}}|}}}
\end{equation*}
The theorem follows. $\hfill \blacksquare$

Actually, $\sum\nolimits_{i \in {{{\mathcal{A}}}_m}} {{2^{|{{\mathcal{I}}}_i^{0} \cap {{\mathcal{A}}}|}}}$ is really the exact value of $|\mathcal{U}_{N,m}|$ in most cases.
\subsection{Complexity}
The complexity of the proposed method to estimate the number of PC-MHW is equal to that of calculating the number of ZCBC-i, for all $i{\rm{\in}}\mathcal{A}_m$. For each $i{\rm{\in}}\mathcal{A}_m$, the complexity of calculating ZCBC-i is $O(\log_2N)$. Thus, the complexity of the proposed estimation is calculated as
\begin{equation}
O(\sum\nolimits_{i \in {{{\mathcal{A}}}_m}} {{2^{|{{\mathcal{I}}}_i^{0} \cap {{\mathcal{A}}}|}}})\leq O(|\mathcal{A}_m|\log_2N)
\end{equation} Unlike the methods in \cite{ProCom} and \cite{enhanceProCom}, the complexity is not much affected by $N$.
\section{Efficient Enumerator Strategy for PC-MHW}
Actually, in previous part we not only provided the upper bound of the number of PC-MHW, but also given that for each $|\mathcal{U}^{(i)}_{N,m}|$, with $i{\rm{\in}}\mathcal{A}_m$. In this part, we will propose an efficient strategy to search specific PC-MHW. We can array $\mathcal{A}_{m}$ and  $\mathcal{I}_i^0$ according to the ascending order, i.e., $\mathcal{A}_{m}=\{\mathcal{A}_{m,1},\mathcal{A}_{m,2},...,\mathcal{A}_{m,|\mathcal{A}_m|}\}$ and $\mathcal{I}_i^0=\{\mathcal{I}^0_{i,1},\mathcal{I}^0_{i,2},...,\mathcal{I}^0_{i,|\mathcal{I}^0_{i}|}\}$. Based on this, we will propose a double multi-level SCLD-based searching strategy which divides $\mathcal{U}_{N,m}^{(i)}$ into $|\mathcal{I}_i^0{\rm{\cap}}\mathcal{A}|{\rm{+}}1$ parts to search, i.e.,
\begin{equation}\label{divideUmi}
\mathcal{U}^{(i)}_{N,m}{\rm{-}}\{\mathbf{g}_N^{(i)}\}{\rm{=}}\bigcup\nolimits_{j \in\mathcal{I}_i^0{\rm{\cap}}\mathcal{A}} {\mathcal{U}_{N,m}^{(i,j) }}
\end{equation}
where ${\mathcal{U}_{N,m}^{(i,j) }}$ is the set of vector $u_1^N$ which satisfies the following two conditions:
\begin{enumerate}
  \item $u_1^{i-1}{\rm{=}}\mathbf{0}_{i-1}$, $u_i{\rm{=}}1$,  $u^{j-1}_{i+1}{\rm{=}}\mathbf{0}_{j-1-i}$, $u_j{\rm{=}}1$, $\mathbf{u}_{\mathcal{A}_{j}}{\rm{\in}}\mathcal{B}^{|\mathcal{A}_{j}|}$, $\mathbf{u}_{\mathcal{A}^c_{j}}{\rm{=}}\mathbf{0}_{|\mathcal{A}^c_j|}$.
  \item $w(u_1^N\mathbf{G}_N)=w(\mathbf{g}_N^{(i)})$.
\end{enumerate}
where $\mathcal{A}_{j}{\rm{=}}\mathcal{A}{\rm{\cap}}\left[\kern-0.15em\left[ {j{\rm{+}}1,N}
 \right]\kern-0.15em\right]$ and $\mathcal{A}^c_{j}{\rm{=}}\mathcal{A}^c{\rm{\cap}}\left[\kern-0.15em\left[ {j{\rm{+}}1,N}
 \right]\kern-0.15em\right]$. Note that when  $j{\rm{=}}i{\rm{+}}1$, $u^{j-1}_{i+1}$ is null vector.
 This strategy can further reduce the list size, so as to reduce the memory space and the sorting complexity required for searching.  In Algorithm 1, we elaborate the process of this strategy. We can find that in the proposed searching strategy, the maximum required list size is $2^{|\mathcal{I}_i^0\cap\mathcal{A}|-1}$, which is less than half of that required for method in \cite{CRC_design} ($|\mathcal{U}_{N,m}^{(i)}|$) and the method in \cite{searchMHWD} ($|\mathcal{U}_{N,m}|{\rm{+}}1$). It is worth noting that the methods in \cite{CRC_design} and \cite{searchMHWD} cannot estimate the number of PC-MHW, in practice the used list size should be set larger than the actually required value.
\begin{algorithm}[!htb]
\caption{Double Multi-level SCLD-based Enumerator for PC-MHW}
\KwIn
{
$y_1^N{\rm{=}}\mathbf{1}_N$, $\mathcal{A}_m$
 }
\KwOut {
$\mathcal{U}_{N,m}$\\}
\For{$k  = 1,2,\cdots,|\mathcal{A}_m|$}
{ $i \Leftarrow \mathcal{A}_{m,k}$\\
 Obtain $\mathcal{I}_i^0$ by \emph{Definition 12};\\
 cnt $\Leftarrow$ 0; \\
$\mathcal{U}_t\Leftarrow \phi$; $\backslash\backslash$ A temporary set to reserve searched paths

\For{$j  = 1,2,\cdots,|\mathcal{I}_i^0|$}
{\If{$\mathcal{I}_{i,j}^0{\rm{\in}}\mathcal{A}$}
       {
       cnt $\Leftarrow$ cnt$+1$;\\
Using SCLD to obtain ${\mathcal{U}_{N,m}^{(i,\mathcal{I}_{i,j}^0) }}$ with list size $L{\rm{=}}2^{|\mathcal{I}_i^0\cap\mathcal{A}|-{\rm{cnt}}}$ under 2 conditions:\\
1. $y_1^N{\rm{=}}\mathbf{1}_N$; 2. $[\hat u_1^{i-1}{\rm{=}}\mathbf{0}_{i-1},\hat u_i{\rm{=}}1,\hat u^{\mathcal{I}_{i,j}^0-1}_{i+1}{\rm{=}}\mathbf{0}_{\mathcal{I}_{i,j}^0-1-i},\hat u_{\mathcal{I}_{i,j}^0}{\rm{=}}1]$.\\
 $\mathcal{U}_t=\mathcal{U}_t\cup{\mathcal{U}_{N,m}^{(i,\mathcal{I}_{i,j}^0) }}$;
}
}
$\mathcal{U}^{(i)}_{N,m}=\mathcal{U}_t\cup\{\mathbf{g}_N^{(i)}\}$;
}
$\mathcal{U}_{N,m}{\rm{=}}\bigcup\nolimits_{i \in\mathcal{A}_m} {\mathcal{U}_{N,m}^{(i) }}$;\\
\textbf{Return} $\mathcal{U}_{N,m}$;
\end{algorithm}
\section{Simulation Results}
\subsection{Accuracy of Estimation}
\begin{figure}
\centering{\includegraphics[height=5cm]{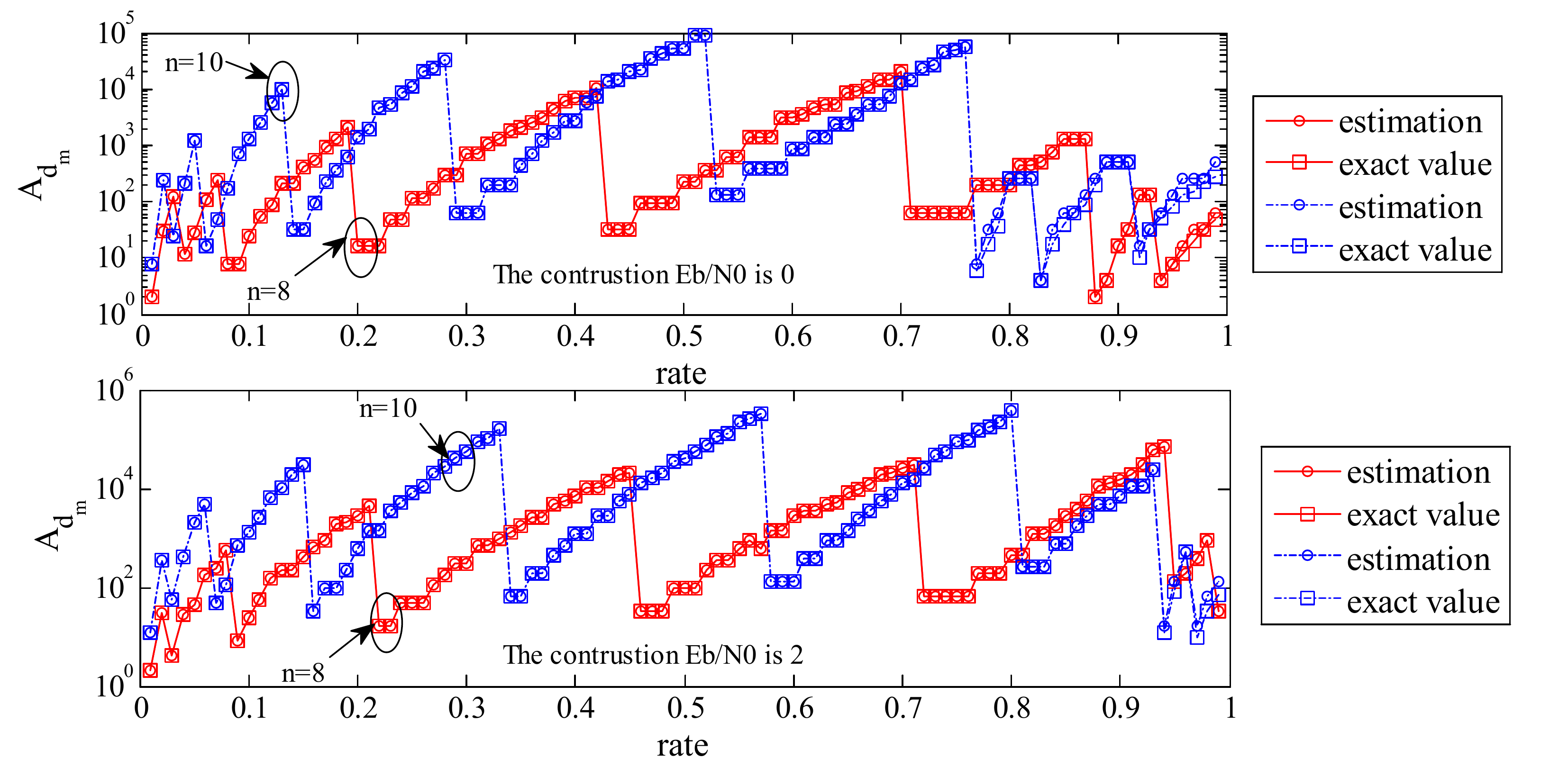}}
\caption{Number of PC-MHW obtained by exhaustive searching and the
proposed evaluating method when using GA algorithm \cite{GA} for construction.}
\end{figure}
\begin{figure}
\centering{\includegraphics[height=5cm]{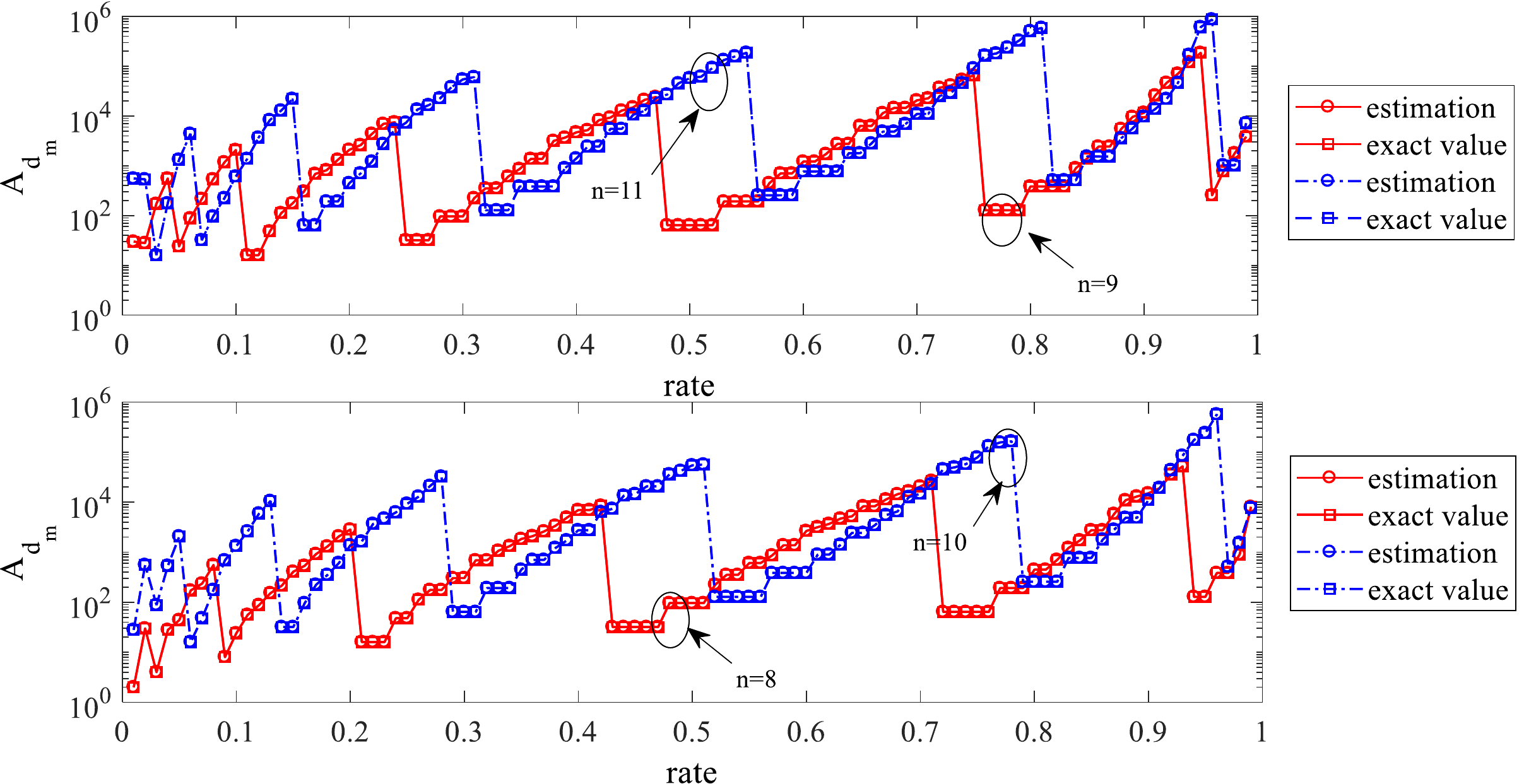}}
\caption{Number of PC-MHW obtained by exhaustive searching and the
proposed evaluating method when using PW algorithm \cite{PW} for construction.}
\end{figure}
Actually, we can directly take the upper bound, i.e., $\sum\nolimits_{i \in {{{\mathcal{A}}}_m}} {{2^{|{{\mathcal{I}}}_i^{0} \cap {{\mathcal{A}}}|}}}$, as the estimated value of the number of PC-MHW. In Fig.2-3, we compare this estimated value between the exact value $|{{{{\mathcal{U}}}_{N,m}}}|$ which comes from exhaustive searching.  We offer the fine-granularity simulation for 99 code rates that arrange from 0.01 to 0.99 with 0.01 step. In Fig.2, the polar codes are constructed by gaussian approximation (GA) algorithm \cite{GA} and the constructing ${{{E_b}} \mathord{\left/{\vphantom {{{E_b}} {{N_0}}}} \right.\kern-\nulldelimiterspace} {{N_0}}}$ is set 0 dB and 2 dB. The code length 1024 and 256 are considered. We can find that for almost all the rates expect some high ones, the evaluated value equals the exact one. In Fig.3, the constructing method is changed to polarization weight (PW) algorithm \cite{PW}. We can find that for all the rates the evaluated value is equal to the exact one, irrespective of the code length.

In Table I,  we also compare the proposed estimation, i.e., $\sum\nolimits_{i \in {{{\mathcal{A}}}_m}} {{2^{|{{\mathcal{I}}}_i^{0} \cap {{\mathcal{A}}}|}}}$, with those proposed in \cite{ProCom} \cite{enhanceProCom}. We consider the code length of 128 and 256. For each length, we adopt 9 code rates (from 0.1 to 0.9 with step 0.1). The estimated number of PC-MHW is denoted as $|\widehat{{{{{\mathcal{U}}}_{N,m}}}}|$. The value for each scheme is calculated by $|\ln(|\widehat{{{{{\mathcal{U}}}_{N,m}}}}|/|{{{{\mathcal{U}}}_{N,m}}}|)|$. The polar codes are designed by GA algorithm with $\sigma^2{\rm{=}}0.6309$. It can be find that the proposed algorithm (denoted by P.) can evaluate the exact value in the entire rate range and is obviously more accurate than the both existing methods.
\begin{table}[tbp]
\centering
\caption{Evaluation for the number of PC-MHW based on the proposed algorithm, the schemes in \cite{ProCom} and \cite{enhanceProCom}.}
\begin{tabular}{l|c|c|c|c|c|c|c|c|c}  %
\hline
Scheme &0.1 &0.2 &0.3 &0.4 &0.5 &0.6 &0.7 &0.8 &0.9\\ \hline\hline  
128,\cite{ProCom}&1.2 &3.1&2.3 &4.7&6.8&8.4&13.7&15.4&20.7 \\ \hline
128,\cite{enhanceProCom} &0 &0&0.2&1.2&0&0&0&0&0 \\ \hline
128,P. &0 &0&0 &0&0&0&0&0&0\\ \hline
256,\cite{ProCom} &2.3 &3.1&5.2 &7.1&12.7&18.4&29.2&38.4&48.1\\ \hline
256,\cite{enhanceProCom} &2.1 &0.8&0.1 &0.5&0.1&0&0&0&0 \\ \hline
256,P. &0 &0&0 &0&0&0&0&0&0\\ \hline
\end{tabular}
\end{table}

\begin{figure}
\centering{\includegraphics[height=4.8cm]{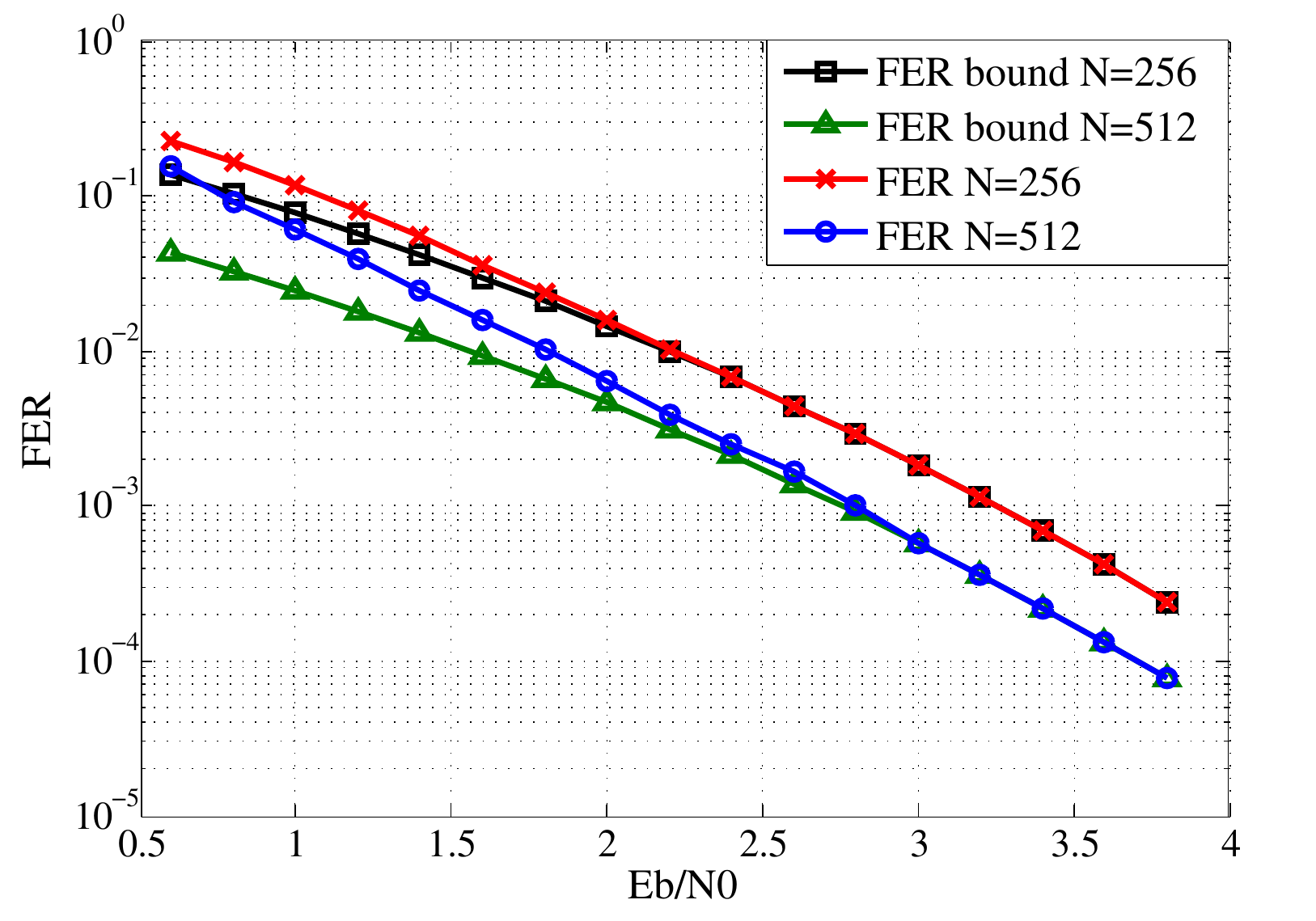}}
\caption{The FER and FER bounds give in (\ref{proposedbound}) of polar code with
$N=\{256,512\}$, code rate $R=0.3$ and list size $L=8$.}
\end{figure}
\begin{figure}
\centering{\includegraphics[height=5cm]{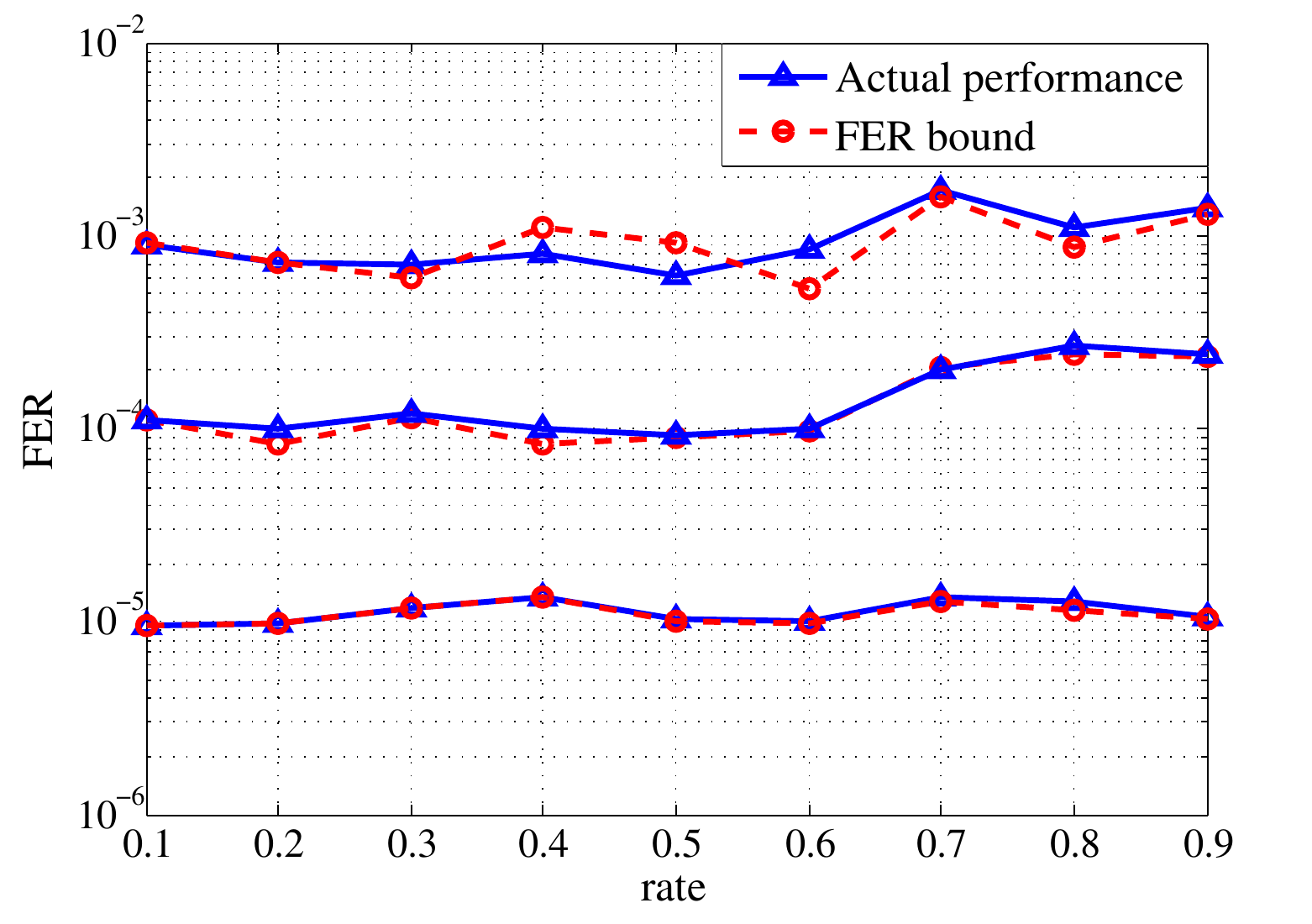}}
\caption{Comparison of the actual performance of SCLD with the bound given in (\ref{proposedbound}) on the order of $10^{-3}$, $10^{-4}$ and $10^{-5}$ FER. The polar code with $N=1024$ is decoded by SCLD with $L=8$.}
\end{figure}
\subsection{Estimation for Performance of SCLD}
When polar codewords are transmitted through AWGN channel, it can approach the ML performance under SCLD. Thus, the performance can be upper bounded by union bound \cite{MLbound}
\begin{equation}
{P_{un}} \leq \sum\limits_d {{A_d}Q\left( {\frac{{\sqrt d }}{\sigma }} \right)}
\end{equation}
where $A_d$ is the number of polar codewords with weight $d$.
At high SNR, the ML performance is dominated by the item with MHW.
Based on \emph{Theorem 17}, at high SNR, the frame error rate (FER) of polar codes under SCLD can be estimated by
\begin{equation}\label{proposedbound}
{P_{SCL}} \approx{{A_{d_m}}Q\left( {\frac{{\sqrt {d_m} }}{\sigma }} \right)}\leq  \sum\nolimits_{i \in {{{\mathcal{A}}}_m}} {{2^{|{{\mathcal{I}}}_i^{0} \cap {{\mathcal{A}}}|}}}\cdot Q\left( {\frac{{\sqrt {d_m} }}{\sigma }} \right)
\end{equation}

In Fig.4, we give the comparison between the actual performance of SCLD with the proposed performance bound given in (\ref{proposedbound}). The considered code lengths are in $\{256,512\}$. For each code length, we conduct the simulation under code rate $R=0.3$. The polar codes are constructed by PW algorithm. The list size of SCLD is $L{\rm{=}}8$. It can be find that the proposed bound is very closed to the performance of SCLD at high SNR. Unlike union bound, which is an upper bound for ML performance, the proposed bound seems a lower bound because we only consider the item associated with PC-MHW. In Fig.5, we give a fine-granularity simulation for comparing the actual performance of SCLD with bound given in (\ref{proposedbound}) on the order of $10^{-3}$, $10^{-4}$ and $10^{-5}$ FER. For polar code with $N{\rm{=}}1024$, constructed by PW algorithm, we consider 9 code rates  (from 0.1 to 0.9 with step 0.1). The list size of SCLD is 8. It can be seen that in most of the cases, the proposed bound can estimate the actual performance well.
\subsection{Comparison with Existing Methods}
In Table II, we compare the proposed enumeration (including PC-MHW and its number) with the existing methods. Among all the methods for evaluating $|\mathcal{U}_{N,m}|$, the complexity of the proposed method is minimum. Moreover, it is suitable for all the code rates. As for enumerating PC-MHW, \cite{ProCom} and \cite{enhanceProCom} do not provide  solution. The proposed method needs less than half of the list size required in \cite{searchMHWD} and \cite{CRC_design}.
\begin{table}[tbp]
\centering
\caption{Comparison of the proposed strategy with the existing schemes.}
\begin{tabular}{c|c|c|c|c|c}  %

\hline
Scheme &\cite{searchMHWD} &\cite{CRC_design} &\cite{ProCom} &\cite{enhanceProCom} &Proposed \\ \hline\hline  
complexity for evaluating $|\mathcal{U}_{N,m}|$&$O(|\mathcal{U}_{N,m}|N\log_2N)$ &$O(|\mathcal{U}_{N,m}|N\log_2N)$&$O(N^5)$ &$O(N^3)$&$O(|\mathcal{A}_m|\log_2N)$ \\ \hline
 suitable code rates for evaluating &all&all&low&high&all\\ \hline
maximum $L$ to search PC-MHW &$|\mathcal{U}_{N,m}|$ &$|\mathcal{U}_{N,m}^{(i)}|$&$\times$&$\times$&$2^{|\mathcal{I}_i^0\cap\mathcal{A}|-1}$ \\ \hline
\end{tabular}
\end{table}
\section{Conclusion}
In this paper, we proposed an efficient method to enumerate the PC-MHW and its number. First, we revealed how the PC-MHW can be enumerated by the existing multi-level SCLD-based schemes \cite{CRC_design}, with $y_1^N{\rm{=}}\mathbf{1}_N$, and obtained the necessary and sufficient condition for a searched path $\hat u_N[l]_1^N$ in $\mathcal{U}_{N,m}^{(i)}$, that is, its RDS equals to $\{i\}$. Subsequently, we introduced a concept of ZCBC-i which will be used to given an upper bound for $|\mathcal{U}^{(i)}_{N,m}|$ and further to derive the upper bound of the number of PC-MHW, i.e., $|\sum\nolimits_{i \in {\mathcal{A}_m}} {\mathcal{U}_{N,m}^{(i)}}|{\rm{\leq}}\sum\nolimits_{i \in {{{\mathcal{A}}}_m}} {{2^{|{{\mathcal{I}}}_i^{0} \cap {{\mathcal{A}}}|}}}$. Guided by the previous analysis, we proposed a double multi-level SCLD-based searching strategy to enumerate all the PC-MHW, which can further divide ${\mathcal{U}_{N,m}^{(i)}}$ into several subsets to search. The maximum required list size for the proposed searching strategy is much less than that for the existing methods, so as to reduce the complexity and memory space required for searching.
\ifCLASSOPTIONcaptionsoff
  \newpage
\fi

\appendices
\section{}
\emph{Proof of Theorem 7}: We first consider the case of $\hat u_N[l]_1^N{\rm{=}}\mathbf{g}_N^{(i)}{\rm{\in}}\mathcal{U}_{N,m}^{(i) }$.
 Based on the hard decision rule given in (\ref{HBDSCL}), proving $\overline{\mathcal{A}}_N[l]{\rm{=}}\{i\}$ is equivalently to proving for any $j{\rm{=}}1,2,...,N$, $L_N[l]_j{\rm{\geq}}0$ and $L_N[l]_i{\rm{>}}0$. Note that when decoding LLR is 0, the corresponding decoding bit, whether decoded to be 0 or 1, can be seen to be decided based on the decoding LLR.

Based on the condition: $y_1^N{\rm{=}}\mathbf{1}_N$ and $\hat u_N[l]_1^{i-1}{\rm{=}}\mathbf{0}_{i-1}$, $L_N[l]_i{\rm{>}}0$ can be easily proved. Then, we will use contradiction method to prove that none of the decoding LLRs of $\hat u_N[l]_1^N$ is negative.

We assume that $\mathcal{P}_{nLLR}$ is the set of positions at which the decoding LLR of $\hat u_N[l]_1^{N}$ is negative. We will determine  $\mathcal{P}_{nLLR}$ by retrace the searching process of $\hat u_N[l]_1^{N}$. Actually, we only need to  discuss the existence of the first position where the negative decoding LLR appears, i.e., the minimum value in $\mathcal{P}_{nLLR}$, denoted by $j_0$. If the conclusion of the retracing is that $j_0$ does not exist, then $\mathcal{P}_{nLLR}{\rm{=}}\phi$, and it follows that $\overline{\mathcal{A}}_N[l]{\rm{=}}\{i\}$. Since $\hat u_N[l]_i{\rm{=}}1$ is the fundamental reason to cause $L_N[l]_{j_0}{\rm{<}}0$, we can easily obtain that ${j_0}{\rm{>}}i$.

Now we can use RSCLD to explore how these negative LLRs were generated in the searching. On the code tree, to obtain ${{\alpha[V_0^{(j_0)}]}}{\rm{<}}0$, the LLR vector of all the GAN of $V_0^{(j_0)}$ should be calculated in advance (from \emph{Property 1}). Thus, at least one of the GAN of $V_0^{(j_0)}$ has negative LLR in its LLR vector. We can write $\mathcal{G}[V_0^{(j_0)}]$ as follows
\begin{equation}
\mathcal{G}[V_0^{(j_0)}]=\{V_0^{(j_0)},V_1^{(j_1)},...,V_N^{(j_N)}\}
\end{equation}
In $\mathcal{G}[V_0^{(j_0)}]$, we assume that $V_f^{(j_f)}$ is the first node whose LLR vector has negative LLR in the searching process, with $0{\rm{\leq}}f{\rm{<}}N$. Since the LLR message is updated from the root node to the leaf node, this means that $\alpha[V_f^{(j_f)}]$ has negative LLR and for any $k$ that satisfies $f{\rm{<}}k{\rm{\leq}}N$, $\alpha[V_k^{(j_k)}]$ has no negative element. Based on this assumption, the LLR vector of the parent node of $V_f^{(j_f)}$, i.e., $\alpha[V_{f+1}^{(\left\lceil {{{{j_f}} \mathord{\left/
 {\vphantom {{{j_f}} 2}} \right.
 \kern-\nulldelimiterspace} 2}} \right\rceil )}]$, has no negative element. This implies that the negative LLR in $\alpha[V_f^{(j_f)}]$ can not be introduced by (\ref{LLRnobit}). In other words, $\alpha[V_f^{(j_f)}]$ can only be calculated by (\ref{LLRbit}) in which $\alpha[V_{f+1}^{(\left\lceil {{{{j_f}} \mathord{\left/
 {\vphantom {{{j_f}} 2}} \right.
 \kern-\nulldelimiterspace} 2}} \right\rceil )}]$ and $\beta[V_f^{(j_f-1)}]$ will be used. Concretely,
 \begin{equation}\label{theoLLRbit}
\alpha{[ {{{V}}_{f}^{\left( {j_f} \right)}} ]_i} {\rm{=}} ( {1 {\rm{-}} 2\beta{{[ {{{V}}_{f}^{\left( {j_f - 1} \right)}} ]}_i}} )\alpha[V_{f+1}^{(\left\lceil {{{{j_f}} \mathord{\left/
 {\vphantom {{{j_f}} 2}} \right.
 \kern-\nulldelimiterspace} 2}} \right\rceil )}]_i {\rm{+}} \alpha[V_{f+1}^{(\left\lceil {{{{j_f}} \mathord{\left/
 {\vphantom {{{j_f}} 2}} \right.
 \kern-\nulldelimiterspace} 2}} \right\rceil )}]_{i+2^f}
\end{equation}
 with $i{\rm{\in}}\left[\kern-0.25em\left[ {1,2^f}
 \right]\kern-0.25em\right]$.
Since $\alpha[V_{f+1}^{(\left\lceil {{{{j_f}} \mathord{\left/
 {\vphantom {{{j_f}} 2}} \right.
 \kern-\nulldelimiterspace} 2}} \right\rceil )}]$ has no negative element,  a necessary condition of generating negative LLR in $\alpha[V_f^{(j_f)}]$   should be
\begin{equation}\label{proveleftchild}
\beta[V_f^{(j_f-1)}]\neq\mathbf{0}_{2^f}.
\end{equation}
Since $\hat u_N[l]_1^N{\rm{=}}$$[\mathbf{0}_{i-1},1, \mathbf{0}_{N-i}]$, to satisfy (\ref{proveleftchild}), $\beta[V_0^{(i)}]{\rm{=}}\hat u_N[l]_i$ should participate in the calculation of $\beta[V_f^{(j_f-1)}]$ by using (\ref{debit}). From \emph{Property  2},
$V_f^{(j_f-1)}$ is the GAN of $V_0^{(i)}$. Further, since the parent node $V_{f+1}^{(\left\lceil {{{{j_f}} \mathord{\left/
 {\vphantom {{{j_f}} 2}} \right.
 \kern-\nulldelimiterspace} 2}} \right\rceil )}$ is the GAN of $V_f^{(j_f-1)}$ (\emph{Property  3}), according to \emph{Property  4} we can obtain that
 \begin{equation}\label{GANThe2}
 V_{f+1}^{(\left\lceil {{{{j_f}} \mathord{\left/
 {\vphantom {{{j_f}} 2}} \right.
 \kern-\nulldelimiterspace} 2}} \right\rceil )}{\rm{\in}}\mathcal{G}[V_0^{(i)}].
   \end{equation}
This means that $\alpha[V_{f+1}^{(\left\lceil {{{{j_f}} \mathord{\left/
 {\vphantom {{{j_f}} 2}} \right.
 \kern-\nulldelimiterspace} 2}} \right\rceil )}]$ is already updated to obtain $\alpha[V_0^{(i)}]$. Thus, $\alpha[V_{f+1}^{(\left\lceil {{{{j_f}} \mathord{\left/
 {\vphantom {{{j_f}} 2}} \right.
 \kern-\nulldelimiterspace} 2}} \right\rceil )}]$ is calculated at the first decoding phase. From \emph{Lemma 4}, all the LLRs in $\alpha[V_{f+1}^{(\left\lceil {{{{j_f}} \mathord{\left/
 {\vphantom {{{j_f}} 2}} \right.
 \kern-\nulldelimiterspace} 2}} \right\rceil )}]$ should be positive and identical. Using (\ref{theoLLRbit}), for any $i{\rm{\in}}\left[\kern-0.25em\left[ {1,2^f}
 \right]\kern-0.25em\right]$, we have
  \begin{equation}
\alpha{[ {{{V}}_{f}^{\left( {j_f} \right)}} ]_i}{\rm{=}}\left\{ \begin{gathered}
0, \quad\quad\quad\quad\quad\quad\;\;\,\, {\rm{if}}\;\beta{{[ {{{V}}_{f}^{\left( {j_f - 1} \right)}} ]}_i}{\rm{=}}1 \hfill \\
  2\alpha[V_{f+1}^{(\left\lceil {{{{j_f}} \mathord{\left/
 {\vphantom {{{j_f}} 2}} \right.
 \kern-\nulldelimiterspace} 2}} \right\rceil )}]_i,\quad\; \;{\rm{if}}\;\beta{{[ {{{V}}_{f}^{\left( {j_f - 1} \right)}} ]}_i}{\rm{=}}0  \hfill \\
\end{gathered}  \right.
\end{equation}
Thus, $\alpha{[ {{{V}}_{f}^{\left( {j_f} \right)}} ]}$ has no negative LLR. This is contradicted with the assumption. It follows that $j_0$ does not exist and $\mathcal{P}_{nLLR}{\rm{=}}\phi$. Therefore, for searched path $\hat u_N[l]_1^N{\rm{=}}$ $[\mathbf{0}_{i-1}, 1, \mathbf{0}_{N-i}]$, we can obtain that the all its decoding LLRs are non-negative and ${ {\alpha[V_0^{(i)}]} }{\rm{>}}0$. It indicates that $\overline{\mathcal{A}}_N[l]=\{i\}$. Further, using (\ref{PM}), we can obtain that the PM of such searched path is $\left|\alpha[V_0^{(i)}]\right|$.

Next, let us consider the case of $\hat u_N[l]_{1}^N{\rm{\in}}\mathcal{U}_{N,m}^{(i) }{\rm{-}}\{\mathbf{g}_N^{(i)}\}$.
Since $w(\hat u_N[l]_1^N\mathbf{G}_N){\rm{=}}w(\mathbf{g}_N^{(i)})$, from \emph{Theorem 6}, the PM of $\hat u_N[l]_1^N$ is equal to that of the searched path $[\mathbf{0}_{i-1},1,\mathbf{0}_{N-i}]$. Recalling  we already obtained that the PM of the searched path $[\mathbf{0}_{i-1},1,\mathbf{0}_{N-i}]$ is $\left|\alpha[V_0^{(i)}]\right|$ in the above, then based on (\ref{PM}) we have
 \begin{equation}
 \textsf{PM}[l]_N{\rm{=}}{\tiny{\sum\nolimits_{j \in \overline{\mathcal{A}}_N[l]}}} {\left| \alpha[V_0^{(j)}]\right|}{\rm{=}}\left|\alpha[V_0^{(i)}]\right|.
  \end{equation}
 Meanwhile, from \emph{Lemma 5}, $\hat u_N[l]_1^{N}$ has $\overline{\mathcal{A}}_N[l]{\rm{\supseteq}}\left\{ {{i}} \right\}$. Thus, $\overline{\mathcal{A}}_N[l]$ can only be $\left\{ {{i}} \right\}$.

According to the proof of the two cases, the theorem is true for any searched path $\hat u_N[l]_{1}^N{\rm{\in}}\mathcal{U}_{N,m}^{(i) }$.

\section{}
\emph{Proof of Lemma 15:}
Decoding $y_1^N$ is equivalent to decoding a punctured polar code \cite{Puncture}. On the code tree, each stage has $N$ LLRs, even if such LLRs may belong to different nodes. For the $\lambda$-th stage, $\lambda{\rm{\in}}\left[\kern-0.15em\left[ {0,n{\rm{-}}1} \right]\kern-0.15em\right]$, we can array them in one single vector, i.e.,
\begin{equation}
\Gamma_\lambda=[\alpha[V_{\lambda}^{(1)}],\alpha[V_{\lambda}^{(2)}],...,\alpha[V_{\lambda}^{(2^{n-\lambda})}]\,].
\end{equation}
Note that we only care about the locations of the zero-valued LLRs at each stage, i.e., $\mathcal{P}_0[\Gamma_\lambda]$.

This lemma can be proved by mathematical induction. We first consider the $n$-th stage. LLR vector $\alpha[V_n^{(1)}]$ is divided into $2^{n-1}$ combinations to update the $N$ LLRs at the $n{\rm{-}}1$ stage (i.e., $\alpha[V_{n-1}^{(1)}]$ and $\alpha[V_{n-1}^{(2)}]$). The combination is expressed as $(\alpha[V_n^{(1)}]_j,\alpha[V_n^{(1)}]_{j+2^{n-1}})$, $j{\rm{\in}}\left[\kern-0.25em\left[ {1,2^{n-1}} \right]\kern-0.25em\right]$, and its corresponding output LLR combination at the $n{\rm{-}}1$-th stage can be expressed as $(\alpha[V_{n-1}^{(1)}]_j,\alpha[V_{n-1}^{(2)}]_{j})$. From \cite{Puncture}, \emph{in SCD}, we can obtain the following relations:
\begin{equation*}
(\alpha[V_n^{(1)}]_j{\rm{=}}0,\alpha[V_n^{(1)}]_{j+2^{n-1}}{\rm{=}}0){\rm{  \Rightarrow  }}(\alpha[V_{n-1}^{(1)}]_j{\rm{=}}0,\alpha[V_{n-1}^{(2)}]_{j}{\rm{=}}0)
\end{equation*}
\begin{equation*}
(\alpha[V_n^{(1)}]_j{\rm{\neq}}0,\alpha[V_n^{(1)}]_{j+2^{n-1}}{\rm{=}}0){\rm{  \Rightarrow  }}(\alpha[V_{n-1}^{(1)}]_j{\rm{=}}0,\alpha[V_{n-1}^{(2)}]_{j}{\rm{\neq}}0)
\end{equation*}
\begin{equation*}
(\alpha[V_n^{(1)}]_j{\rm{=}}0,\alpha[V_n^{(1)}]_{j+2^{n-1}}{\rm{\neq}}0){\rm{  \Rightarrow  }}(\alpha[V_{n-1}^{(1)}]_j{\rm{=}}0,\alpha[V_{n-1}^{(2)}]_{j}{\rm{\neq}}0)
\end{equation*}
\begin{equation}\label{relation}
(\alpha[V_n^{(1)}]_j{\rm{\neq}}0,\alpha[V_n^{(1)}]_{j+2^{n-1}}{\rm{\neq}}0){\rm{  \Rightarrow  }}(\alpha[V_{n-1}^{(1)}]_j{\rm{\neq}}0,\alpha[V_{n-1}^{(2)}]_{j}{\rm{\neq}}0)
\end{equation}
where the notation  $\Rightarrow$ means generating by (\ref{LLRnobit}) and (\ref{LLRbit}).

Let $c_{1}^{N}{\rm{=}}u_1^N\mathbf{G}_N$ and it can be expressed as
\begin{equation}\label{c}
c_{1}^{N}{\rm{=}}[u_1^{N/2}\mathbf{G}_{N/2}{\rm{\oplus}}u_{N/2+1}^{N}\mathbf{G}_{N/2}, u_{N/2+1}^{N}\mathbf{G}_{N/2}].
\end{equation}
Then, we have
\begin{equation}\label{weight}
w(c_1^{N}){\rm{=}}w(u_1^{N/2}\mathbf{G}_{N/2}){\rm{+}}2w(c_{N/2+1}^{N}){\rm{-}}2v,
\end{equation}
where $v$ is the number of the bits in $u_1^{N/2}\mathbf{G}_{N/2}$ and $c_{N/2+1}^{N}$ equal to 1 at the same position. Obviously, one has
\begin{equation}\label{input}
y_1^N{\rm{=}}a\cdot(\mathbf{1}_N{\rm{-}}c_1^N){\rm{=}}\alpha[V_n^{(1)}]
\end{equation}
If $i{\rm{>}}2^{n-1}$ (or $b_{i-1,n}{\rm{=}}1$), then $u_1^{N/2}\mathbf{G}_{N/2}{\rm{=}}\mathbf{0}$. Using (\ref{c}), we have
\begin{equation}\label{c1}
c_1^{2^{n-1}}{\rm{=}}c^N_{1+2^{n-1}}.
\end{equation}

Substituting (\ref{c1}) into (\ref{input}), we can obtain that in vector $\alpha[V_n^{(1)}]$ there can only exist the combination of $(\alpha[V_n^{(1)}]_j{\rm{=}}0,\,\alpha[V_n^{(1)}]_{j+2^{n-1}}{\rm{=}}0)$, and $(\alpha[V_n^{(1)}]_j{\rm{>}}0,\,\alpha[V_n^{(1)}]_{j+2^{n-1}}{\rm{=}}0)$ or $(\alpha[V_n^{(1)}]_j{\rm{=}}0,$ $\alpha[V_n^{(1)}]_{j+2^{n-1}}{\rm{>}}0)$ does not exist. In this case,  from (\ref{relation}), $\mathcal{P}_0[\Gamma_{n-1}]$ can be obtained by:
\begin{equation}\label{0loc1}
\mathcal{P}_0[\Gamma_{n-1}]=\mathcal{P}_0[\alpha[V_{n}^{(1)}]\,]=\mathcal{P}_1[c_1^N]
\end{equation}

Otherwise, we consider the case of $i{\rm{\leq}}2^{n-1}$ (or $b_{i-1,n}{\rm{=}}0$). Since  $u_1^N{\rm{\in}}\mathcal{U}_{N,m}^{(i)}$, it follows that $w(c_1^{N}){\rm{=}}w(\mathbf{g}_N^{(i)})$. Meanwhile, from \emph{Lemma 2} in \cite{CRC_design}, we have
\begin{equation}
w(u_1^{{N \mathord{\left/
 {\vphantom {N 2}} \right.
 \kern-\nulldelimiterspace} 2}}\mathbf{G}_{{N \mathord{\left/
 {\vphantom {N 2}} \right.
 \kern-\nulldelimiterspace} 2}}){\rm{\geq}}w(\mathbf{g}_N^{(i)}){\rm{=}}w(c_1^N).
 \end{equation}
If $w(u_1^{{N \mathord{\left/
 {\vphantom {N 2}} \right.
 \kern-\nulldelimiterspace} 2}}\mathbf{G}_{{N \mathord{\left/
 {\vphantom {N 2}} \right.
 \kern-\nulldelimiterspace} 2}}){\rm{>}}w(c_1^N)$, (\ref{weight}) implies that $w(c_{N/2+1}^{N}){\rm{<}}v$. This contradicts with the definition of $v$. Therefore, it can only be $w(u_1^{{N \mathord{\left/
 {\vphantom {N 2}} \right.
 \kern-\nulldelimiterspace} 2}}\mathbf{G}_{{N \mathord{\left/
 {\vphantom {N 2}} \right.
 \kern-\nulldelimiterspace} 2}}){\rm{=}}w(c_1^N)$. From (\ref{weight}), we can further obtain that $w(c_{N/2+1}^{N}){\rm{=}}v$. This means that  for any $j{\rm{\in}}\left[\kern-0.15em\left[ {1,2^{n-1}} \right]\kern-0.15em\right]$,  $c_j$ and $c_{j+2^{n-1}}$ can not be 1 simultaneously and the combination $(\alpha[V_n^{(1)}]_j{\rm{=}}0,\alpha[V_n^{(1)}]_{j+2^{n-1}}{\rm{=}}0)$ does not exist. This lead to
 \begin{equation}\label{0loc2}
 \mathcal{P}_0[\Gamma_{n-1}]=\mathcal{P}_1[c_1^{\frac{N}{2}}]\cup\mathcal{P}_1[c_{\frac{N}{2}+1}^N].
 \end{equation}

Then, at the $n{\rm{-}}1$-th stage, the code tree can be divided into two subcode-trees. The first one is rooted at $V_{n-1}^{(1)}$ and the second one is rooted at $V_{n-1}^{(2)}$. From the above, we can obtain that if $b_{i-1,n}{\rm{=}}1$, it has $\mathcal{P}_0[\alpha[V_{n-1}^{(1)}]\,]=\mathcal{P}_0[\alpha[V_{n-1}^{(2)}]\,]$. Since the two subcode-trees performs independently, in their respective following decoding stage, the set of the locations of zero-valued LLRs  will be identical with each other. Meanwhile, if $b_{i-1,n}{\rm{=}}0$, the second subcode-tree can not generate any 0-valued LLR in its descendant nodes.

From the above analysis, for any stage $\lambda{\rm{\in}}\left[\kern-0.15em\left[ {0,n {\rm{-}} 1}
 \right]\kern-0.15em\right]$, we can conclude that
  \begin{equation}\label{theo13subcodetree1}
\mathcal{P}_0[{\Gamma^\vdash_\lambda }] {\rm{=}} \left\{ \begin{gathered}
  \mathcal{P}_0[{\Gamma^\dashv _\lambda }],\quad\; b_{i-1,n}{\rm{=}}1 \hfill \\
  \phi,\,\,\quad\quad\quad b_{i-1,n}{\rm{=}}0\; \hfill\\
\end{gathered}  \right.
\end{equation}
where $\Gamma^\dashv _\lambda$ and ${\Gamma^\vdash_\lambda }$ are the subvectors of  $\Gamma_\lambda$ with its first and last $\frac{N}{2}$ elements, respectively. That's to say, to obtain $\mathcal{P}_0[{\Gamma_{n-2}}]$, it is enough to focus on the first subcode at the $n{\rm{-}}1$ stage, but instead of considering the both subcodes.

To facilitate induction, based on (\ref{0loc1}) and (\ref{0loc2}), we can write $\mathcal{P}_0[{\Gamma_{n-1} }]$ as:
 \begin{equation}\label{facilitate}
\mathcal{P}_0[{\Gamma_{n-1} }] {\rm{=}} \left\{ \begin{gathered}
  \mathcal{P}_1[\,u_1^{N}\mathbf{G}_N],\quad\quad\quad\quad\quad\quad\;\;\, b_{i-1,n}{\rm{=}}1 \hfill \\
  \mathcal{P}_1[\,[u_1^{N-2^{n-1}},\mathbf{0}_{2^{n-1}}]\mathbf{G}_N],\quad\;\, b_{i-1,n}{\rm{=}}0\; \hfill\\
\end{gathered}  \right.
\end{equation}

Similarly, for the $\lambda$-th stage, $\lambda{\rm{\in}}\left[\kern-0.15em\left[ {1,n {\rm{-}} 2}
 \right]\kern-0.15em\right]$, we assume that
 \begin{equation}\label{theo13assump}
\mathcal{P}_0[{\Gamma_{\lambda} }] {\rm{=}} \left\{ \begin{gathered}
  \mathcal{P}_1[\,u_1^{N}\mathbf{G}_N],\;\;\;\;\,\quad\quad\;\quad [b_{i-1,\lambda+1},...,b_{i-1,n}]{\rm{=}}\mathbf{1}_{n-\lambda} \hfill \\
  \mathcal{P}_1[\,[u_1^{N-k_\lambda},\mathbf{0}_{k_\lambda}]\mathbf{G}_N],\quad {\rm{otherwise}}\; \hfill\\
\end{gathered}  \right.
\end{equation}
where ${k_\lambda } {\rm{=}} \sum\nolimits_{j = \lambda  + 1}^n {\left( {1 {\rm{-}}{b_{i - 1,j}}} \right)} {2^{j - 1}}$.

 To calculate $\mathcal{P}_0[{\Gamma_{\lambda-1} }]$, we can also divide the code-tree nodes from stage 0 to stage $\lambda$ into $2^{n-\lambda}$ subcode-trees. From (\ref{theo13subcodetree1}), some of the LLR vectors of the $2^{n-\lambda}$ root nodes have no zero-valued LLR and thus their descendant nodes also have no 0-valued LLR. We can determine such root nodes to simplify the computation. Before that, we first define $\mathcal{S}_{n}^{j}$ as the set of integer $i{\rm{\in}}\left[\kern-0.15em\left[ {0,n {\rm{-}} 1}
 \right]\kern-0.15em\right]$ whose binary expansion satisfies $b_{i,j}=1$, with $j{\rm{\in}}\left[\kern-0.15em\left[ {1,n {\rm{-}} 1}
 \right]\kern-0.15em\right]$. Obviously, $|\mathcal{S}_{n}^{j}|=2^{n-1}$.

\emph{Example 5: $\mathcal{S}_{1}^{1}{\rm{=}}\{1\}$, $\mathcal{S}_{2}^{1}{\rm{=}}\{1,3\}$, $\mathcal{S}_{2}^{2}{\rm{=}}\{2,3\}$, $\mathcal{S}_{3}^{1}{\rm{=}}\{1,3,5,7\}$, $\mathcal{S}_{3}^{2}{\rm{=}}\{2,3,6,7\}$, $\mathcal{S}_{3}^{3}{\rm{=}}\{4,5,6,7\}$.}

Using equation (\ref{theo13subcodetree1}) recursively from stage $n-1$ to stage $\lambda$, for any $k{\rm{\in}}\left[\kern-0.25em\left[ {0,2^{n-\lambda} {\rm{-}} 1}
 \right]\kern-0.25em\right]$ we have
 \begin{equation}\label{lambdaset}
\mathcal{P}_0[\alpha[V^{(k+1)}_{\lambda}]\,] = \left\{ \begin{gathered}
  \phi,\quad\quad\quad\quad\;\;\;{\rm{if}}\;k{\rm{\in}}\mathcal{S},  \hfill \\
  \mathcal{P}_0[\alpha[V^{(1)}_{\lambda}]\,],\;\,\,{\rm{otherwise}}, \hfill \\
\end{gathered}  \right.
\end{equation}
where $ \mathcal{S}=\bigcup\nolimits_{j \in{\mathcal{P}_0}[\mathbf{b}_n(i-1)_{\lambda+1}^n]} {{{\mathcal{S}}}_{n - \lambda }^{j}}$. Note that if  ${\mathcal{P}_0}[\mathbf{b}_n(i-1)_{\lambda+1}^n]=\phi$, then $ \mathcal{S}=\phi$.

 Obviously, since $0{\rm{\notin }}\bigcup\nolimits_{j =1}^{n-\lambda} {{{\mathcal{S}}}_{n - \lambda }^{j}}$, then $\mathcal{P}_0[\alpha[V^{(1)}_{\lambda}]\,]$ is bound to a non-empty set. That's to say, for each subcode-tree divided at the $\lambda$-th stage, the location set of zero-valued LLR in its input vector can only be null set or equal to $\mathcal{P}_0[\alpha[V^{(1)}_{\lambda}]\,]$.

Actually, to obtain $\mathcal{P}_0[{\Gamma_{\lambda-1} }]$, we do not have to calculate $\mathcal{P}_0[\alpha[V^{(1)}_{\lambda}]\,]$. Instead, we just need to calculate in any subcode-tree (divided at stage $\lambda$) whose LLR vector of its root node has 0-valued LLR.

From (\ref{theo13assump}), if $[b_{i-1,\lambda+1},...,b_{i-1,n}]{\rm{=}}\mathbf{1}_{n-\lambda}$, then $k_\lambda{\rm{=}}0$, and thus
$u_1^N{\rm{=}}[\mathbf{0}_{N-2^\lambda},u^{N}_{N-2^\lambda+1}]$, where $u^{N}_{N-2^\lambda+1}{\rm{\neq}}\mathbf{0}_{2^\lambda}$; if $[b_{i-1,\lambda+1},...,b_{i-1,n}]{\rm{\neq}}\mathbf{1}_{n-\lambda}$, then $[u_1^{N-k_\lambda},\mathbf{0}_{k_\lambda}]{\rm{=}}[\mathbf{0}_{N-k_\lambda-2^\lambda},u^{N-k_\lambda}_{N-k_\lambda-2^\lambda+1}{\rm{\neq}}\mathbf{0}_{2^\lambda},\mathbf{0}_{k_\lambda}]$.

Based on this, for the subcode-tree rooted at $V_\lambda^{(\frac{N-k_\lambda}{2^\lambda})}$ (whose leaf nodes are $V_0^{(N-k_\lambda-2^\lambda+1)}$, $V_0^{(N-k_\lambda-2^\lambda+2)}$,..., $V_0^{(N-k_\lambda)}$), we can obtain the following equation
 \begin{equation}\label{subcodeklambda}
 \mathcal{P}_0[\alpha[V_\lambda^{(\frac{N-k_\lambda}{2^\lambda})}]\,]= \mathcal{P}_1[u^{N-k_\lambda}_{N-k_\lambda-2^\lambda+1}\mathbf{G}_{2^\lambda}]
  \end{equation}
  It can be easily obtained that $N{\rm{-}}k_\lambda{\rm{\geq}}i{\rm{=}}N{\rm{-}}k_0{\rm{\geq}}N{\rm{-}}k_\lambda{\rm{-}}2^\lambda{\rm{+}}1$. From $u_i{\rm{=}}1$, one has $\mathcal{P}_0[\alpha[V_\lambda^{(\frac{N-k_\lambda}{2^\lambda})}]\,]\neq\phi$. Thus, (\ref{lambdaset}) can be rewritten  as
  \begin{equation}\label{lambdaset1}
\mathcal{P}_0[\alpha[V^{(k+1)}_{\lambda}]\,] = \left\{ \begin{gathered}
  \phi,\quad\quad\quad\quad\quad\quad\;\;\,k{\rm{\in}}\mathcal{S},  \hfill \\
  \mathcal{P}_0[\alpha[V_\lambda^{(\frac{N-k_\lambda}{2^\lambda})}]\,],\;\;\,\,{k{\rm{\in}}\left[\kern-0.22em\left[ {0,2^{n-\lambda} {\rm{-}} 1}
 \right]\kern-0.22em\right]{\rm{-}}\mathcal{S}}, \hfill \\
\end{gathered}  \right.
\end{equation}

Since   $\alpha[V_\lambda^{(\frac{N-k_\lambda}{2^\lambda})}]$  generates $[\alpha[V_{\lambda-1}^{(\frac{N-k_\lambda}{2^{\lambda-1}}-1)}],\alpha[V_{\lambda-1}^{(\frac{N-k_\lambda}{2^{\lambda-1}})}]\,]$, (\ref{lambdaset1}) implies that we can determine $\mathcal{P}_0[\Gamma_{\lambda-1}]$ by  calculating $\mathcal{P}_0[\,[\alpha[V_{\lambda-1}^{(\frac{N-k_\lambda}{2^{\lambda-1}}-1)}],\alpha[V_{\lambda-1}^{(\frac{N-k_\lambda}{2^{\lambda-1}})}]\,]\,]$ first.

Note that $u_i$ is the $(i-N+k_\lambda+2^\lambda)$-th bit in $u^{N-k_\lambda}_{N-k_\lambda-2^\lambda+1}$. From \emph{Property  2}, we have $V_\lambda^{(\frac{N-k_\lambda}{2^\lambda})}{\rm{\in}}\mathcal{G}[V_0^{(i)}]$. Meanwhile, since $u_1^N{\rm{\in}}\mathcal{U}_{N,m}^{(i)}$, we can obtain the following equations
\begin{equation}
\begin{gathered}
\begin{aligned}
w(u^{N-k_\lambda}_{N-k_\lambda-2^\lambda+1}\mathbf{G}_{2^\lambda})
= w(\beta[V_\lambda^{(\frac{N-k_\lambda}{2^\lambda})}])
&\mathop  = \limits^{\left( a \right)} w(\mathbf{g}_N^{(i)})/2^{|\mathcal{P}_1[\mathbf{b}_n(i-1)^n_{\lambda+1}]|}\\
\mathop  = \limits^{\left( b \right)}w(\mathbf{g}_{N}^{(i)}(\lambda))
&\mathop  = \limits^{\left( c \right)}w(\mathbf{g}_{2^\lambda}^{(i-N+k_\lambda+2^\lambda)})
\end{aligned}
\end{gathered}
\end{equation}
where step (a) is obtained by \emph{Lemma 13}, step (b) is based on \emph{Lemma 14} and step (c) is based on the structure of $\mathbf{G}_N$. Thus, we can obtain that $u^{N-k_\lambda}_{N-k_\lambda-2^\lambda+1}{\rm{\in}}\mathcal{U}_{2^\lambda,m}^{(i-N+k_\lambda+{2^\lambda})}$. Under this condition, we can refer to the method of calculating $\mathcal{P}_0[\Gamma_{n-1}]$ and obtain the following equation
\begin{equation}
\begin{gathered}\label{lambda_1}
\begin{aligned}
\mathcal{P}_0[\,[\alpha[V_{\lambda-1}^{(\frac{N-k_\lambda}{2^{\lambda-1}}-1)}],\alpha[V_{\lambda-1}^{(\frac{N-k_\lambda}{2^{\lambda-1}})}]\,]\,]
=\left\{ \begin{gathered}
  \mathcal{P}_1[u^{N-k_\lambda}_{N-k_\lambda-2^\lambda+1}\mathbf{G}_{2^\lambda}],\quad\quad \quad\quad \; i{\rm{-}}N{\rm{+}}k_\lambda{\rm{+}}2^\lambda{\rm{\geq}}{2^{\lambda-1}} \hfill \\
  \mathcal{P}_1[\,[u^{N-k_\lambda-2^{\lambda-1}}_{N-k_\lambda-2^\lambda+1},\mathbf{0}_{2^{\lambda-1}}]\mathbf{G}_{2^\lambda}],\,\;\, \; i{\rm{-}}N{\rm{+}}k_\lambda{\rm{+}}2^\lambda{\rm{<}}{2^{\lambda-1}}\; \hfill\\
\end{gathered}  \right.
\end{aligned}
\end{gathered}
\end{equation}
Further, using the fact that $i{\rm{=}}N{\rm{-}}k_0$, it can be easily verified that
\begin{equation*}
i{\rm{-}}N{\rm{+}}k_\lambda{\rm{+}}2^\lambda{\rm{\geq}}{2^{\lambda-1}}\Leftrightarrow b_{i-1,\lambda}{\rm{=}}1,\quad
i{\rm{-}}N{\rm{+}}k_\lambda{\rm{+}}2^\lambda{\rm{<}}{2^{\lambda-1}}\Leftrightarrow b_{i-1,\lambda}{\rm{=}}0
\end{equation*}
From (\ref{lambdaset1}) and (\ref{lambda_1}), we can further obtain that
 \begin{equation*}
\mathcal{P}_0[{\Gamma_{\lambda-1} }] {\rm{=}} \left\{ \begin{gathered}
  \mathcal{P}_1[\,u_1^{N}\mathbf{G}_N],\;\;\;\, [b_{i-1,\lambda},...,b_{i-1,n}]{\rm{=}}\mathbf{1}_{n-\lambda+1} \hfill \\
  \mathcal{P}_1[\,[u_1^{N-k_{\lambda-1}},\mathbf{0}_{k_{\lambda-1}}]\mathbf{G}_N],\quad {\rm{otherwise}}\; \hfill\\
\end{gathered}  \right.
\end{equation*}
By simply induction, at stage 0 we have
 \begin{equation*}
\mathcal{P}_0[{\Gamma_{0} }] {\rm{=}} \left\{ \begin{gathered}
  \mathcal{P}_1[\,u_1^N\mathbf{G}_N]{\rm{=}}\mathcal{P}_1[\,\mathbf{g}_N^{(N)}],\;\;\;\, [b_{i-1,1},...,b_{i-1,n}]{\rm{=}}\mathbf{1}_{n} \hfill \\
  \mathcal{P}_1[\,[u_1^{N-k_{0}},\mathbf{0}_{k_{0}}]\mathbf{G}_N],\quad {\rm{otherwise}}\; \hfill\\
\end{gathered}  \right.
\end{equation*}
Since $i{\rm{=}}N{\rm{-}}k_{0}$ and $u_1^N{\rm{\in}}\mathcal{U}_{N,m}^{(i)}$, then $\mathcal{P}_0[{\Gamma_{0} }]= \mathcal{P}_1[\mathbf{g}_N^{(i)}]$.
The lemma is proved.

\end{document}